\documentclass[11pt,a4paper]{article}
\usepackage[margin=1in]{geometry}

\usepackage{amssymb}
\usepackage{selinput}
\usepackage{url}
\usepackage{amsmath}
\usepackage{ifthen}
\usepackage{color}
\usepackage{graphicx}
\usepackage{multirow}
\usepackage{authblk}
\usepackage{hyperref}
\usepackage{natbib}
\usepackage[table]{xcolor}
\usepackage{booktabs}
\DeclareRobustCommand{\old}[1]{}

\begin{document}

\title{Robust Implicit Neural Representation for Multiparameter Elastic Full-Waveform Inversion}

\renewcommand{\thefootnote}{\fnsymbol{footnote}}

\author[1]{Chao Li}
\author[1]{Yangkang Chen}
\affil[1]{The University of Texas at Austin}

\maketitle
\abstract[Abstract]
{Multiparameter elastic full-waveform inversion (EFWI) provides a powerful means of reconstructing subsurface elastic properties, but its practical application remains challenging because of strong nonlinearity, parameter crosstalk, and the different sensitivities of seismic data to individual elastic parameters. We develop a robust implicit neural representation (INR) framework for multiparameter EFWI, using continuous neural representations to parameterize the subsurface model and provide an effective model-space constraint during inversion. Within this framework, we systematically investigate different elastic parameterizations and demonstrate that parameterization plays a critical role in controlling parameter coupling and inversion stability. Among the tested formulations, combining P-wave velocity (Vp), P-wave impedance (Ip), and the Vs/Vp ratio provides the most robust overall performance, yielding improved parameter separation and more reliable reconstruction of elastic properties. We evaluate the proposed approach on the Marmousi and Overthrust models under increasingly challenging inversion conditions and further apply it to the field Volve ocean-bottom-node dataset. Both synthetic and field-data results demonstrate that the INR-based formulation, together with the Vp–Ip–Vs/Vp parameterization, improves inversion stability, suppresses parameter cross-talk and inversion artifacts, and produces geologically coherent multiparameter models. These results highlight the importance of jointly considering model representation and physical parameterization in multiparameter EFWI and demonstrate the potential of robust INR-based inversion for complex synthetic and field seismic applications.}


\maketitle

\renewcommand\thefootnote{}

\renewcommand\thefootnote{\fnsymbol{footnote}}
\setcounter{footnote}{1}

\section{Introduction}\label{sec1}
Full-waveform inversion (FWI) seeks quantitative subsurface models by minimizing the discrepancy between observed and numerically simulated seismic waveforms \citep{crase1990,vigh2014,hicks2016,barnier2023,Licqawi2024,licdlmfwi2026}. By exploiting the phase, amplitude, and waveform information contained in seismic data, FWI can provide substantially higher spatial resolution than conventional traveltime-based approaches. Since its early development \citep{taran1984}, FWI has evolved into a major tool for high-resolution seismic imaging, enabled by advances in numerical wave-equation solvers, adjoint-state optimization, acquisition technology, and high-performance computing \citep{virieus2009,operto2015}. Most successful applications have nevertheless been based on acoustic or pseudo-acoustic approximations. In geologically complex settings and for multicomponent acquisitions, elastic wave propagation—including P waves, S waves, mode conversions, and their amplitude variations contains additional information that can potentially constrain multiple mechanical properties of the subsurface \citep{licdlmefwi2026}.

\par Multiparameter elastic full-waveform inversion (EFWI) extends conventional FWI by simultaneously estimating several elastic properties. For an isotropic elastic medium, the model can be described using different but mathematically equivalent parameter classes, including the Lamé parameters and density $(\lambda,\mu,\rho)$, P- and S-wave velocities and density $(V_P,V_S,\rho)$, bulk and shear moduli, or combinations involving P- and S-wave impedance. This flexibility is important because mathematically equivalent descriptions of the elastic wave equation do not lead to equivalent inverse problems. The selected parameterization controls the radiation patterns of the associated virtual sources, the relative scaling and sensitivity of model parameters, and the degree to which their Fréchet derivatives overlap. Consequently, parameterization can substantially affect the conditioning, convergence, and robustness of multiparameter EFWI \citep{brossier2009,operto2013,prieux2013,borisov2017,sunh2022}.
\par However, EFWI always suffers from crosstalk. Different elastic parameters may generate partially overlapping perturbations in the seismic wavefield, allowing an observed waveform change caused by one physical property to be incorrectly accommodated by another. The severity of this ambiguity depends on wave mode, scattering angle, frequency content, acquisition aperture, source and receiver components, and the background medium. Radiation-pattern analysis therefore provides an important framework for understanding parameter sensitivity and coupling. Previous studies have investigated parameterizations such as $(\lambda,\mu,\rho)$, $(V_p,V_s,\rho)$, and $(I_p,I_s,\rho)$, demonstrating that their inversion characteristics can differ substantially and that density is generally among the most difficult parameters to be recovered independently \citep{kohn2012,panwy2018}. Studies of multicomponent ocean-bottom data further demonstrated that the choice of parameterization must be considered together with acquisition geometry and the hierarchy of wave modes and data components used during inversion \citep{operto2013,prieux2013}. More generally, numerical experiments show that no parameterization is universally optimal: performance depends on the problem and is governed by interactions among parameter sensitivities, acquisition geometry, optimization, and the underlying geology \citep{Modrak2016}.

\par Consequently, several strategies have been developed to stabilize multiparameter EFWI. These include hierarchical or staged inversion of different parameter classes \citep{kohn2012}, gradient scaling and preconditioning \citep{gebre2026}, approximate-Hessian or Newton-type corrections \citep{wangyw2016,hunagjp2025}, parameter-specific optimization strategies, empirical rock-physics relationships \citep{LI2019106,huqi2021}, structural or facies constraints \citep{kamath2017,aragao2020,singh2021elastic}, physics-guided deep learning based EFWI \citep{dhara2023elastic}, and regularization based on prior geological information \citep{taufik2024,aghazade2026}. Such approaches can substantially reduce parameter contamination, but important limitations remain. In particular, conventional EFWI commonly represents each physical parameter independently on a discrete spatial grid. Every grid point is therefore treated as an independent degree of freedom unless correlations are explicitly introduced through regularization. This high-dimensional representation can facilitate the reconstruction of short-wavelength artifacts and noise and does not intrinsically encode the spatial continuity and multiscale organization expected of geological models. These difficulties become particularly important when several elastic parameters are inverted simultaneously.

\par Implicit neural representations (INRs) provide an alternative way to parameterize spatially varying physical properties \citep{sun2023implicit,chenrh2025,cheny2026,kangb2026}. Rather than storing a model explicitly as values on a grid, an INR represents a continuous field through a coordinate-based neural function, $\mathbf{m}(\mathbf{x}) = f_{\boldsymbol{\theta}}(\gamma(\mathbf{x}))$, where \(\mathbf{x}\) denotes spatial coordinates, \(\gamma(\cdot)\) is an optional coordinate encoding, and the trainable network parameters \(\boldsymbol{\theta}\) implicitly describe the model. Coordinate-based neural representations can compactly represent continuous signals while providing differentiability and controllable spectral characteristics. For inverse problems, this changes the optimization variables from individual model samples to the weights of a function that generates the model. The resulting representation therefore introduces an implicit spatial prior without requiring a pre-existing training dataset. Sun et al. (2023) \citep{sun2023implicit} demonstrated the potential of this idea for seismic inversion by introducing implicit seismic FWI using a deep neural representation of the subsurface velocity model. Their results showed that replacing the conventional grid-based model with a continuous neural representation can improve robustness to initialization and naturally accommodate multiscale model construction. Subsequent studies have extended implicit representations toward multiparameter geophysical inversion and elastic FWI. In particular, Berti et al. (2026) \citep{berit2026} investigated INR-based elastic FWI for near-surface surface-wave applications and demonstrated improved robustness to initialization, while recent developments have explored shared implicit representations of multiple physical properties and meta-learning strategies to accelerate implicit FWI \citep{wangzf2026}. These developments suggest that INR is more than an alternative numerical representation: it can modify the effective model space explored by waveform inversion.

\par However, extending INR from acoustic FWI to multiparameter EFWI is not simply a matter of increasing the number of network outputs. The neural representation does not eliminate the underlying physical ambiguity among elastic parameters. If strongly coupled physical quantities are represented and jointly optimized, the data gradients remain governed by their elastic radiation patterns and acquisition-dependent sensitivities. Conversely, the INR itself couples model updates through shared network parameters and spatial features. Multiparameter implicit EFWI therefore contains two distinct but interacting parameterization problems: how the subsurface is represented mathematically, and which physical variables are represented. The first determines the model space and its implicit regularization, whereas the second determines the sensitivity, scaling, and crosstalk of the elastic inverse problem. We argue that these two aspects should be designed jointly rather than independently.

\par In this study, we develop a robust implicit neural representation framework for multiparameter EFWI and systematically investigate this interaction. We first formulate multiparameter EFWI in an implicit model space, mapping spatial coordinates through a neural representation to multiple elastic-property fields, and optimize the network parameters directly via elastic-wave-equation misfit. We then investigate alternative physical parameterizations through radiation-pattern analysis and controlled inversion experiments. We pay particular attention to the conventional $(Vp,Vs,\rho)$ formulation and alternative velocity-, impedance Ip-, and ratio-based parameterizations. Our experiments show that, for the acquisition geometries and geological settings investigated here, the combination $\left(Vp,\ Ip,\ Vs/Vp\right)$ performs particularly well. In this formulation, Vp primarily describes the compressional kinematics, $Ip=\rho Vp$ provides an impedance-related description of elastic contrasts, and Vs/Vp expresses the relative shear response. Combined with the implicit neural representation, this parameterization substantially reduces detrimental parameter coupling and produces more stable multiparameter reconstruction than conventional alternatives.

\par The main contribution of this work is therefore the joint design of implicit neural model representation and elastic physical parameterization. Rather than treating INR solely as additional regularization for conventional EFWI, we examine how the implicit model space interacts with elastic parameter sensitivity and show that appropriate reparameterization is essential for robust multiparameter INR inversion. We validate the proposed framework using the Marmousi and Overthrust models, which provide distinct structural and elastic complexities, and further evaluate its applicability using the Volve ocean-bottom-node (OBN) field dataset. Across these experiments, the proposed $(V_P,I_P,Vs/Vp)$ implicit formulation improves parameter separation, suppresses crosstalk and inversion artifacts, and produces stable, geologically coherent multiparameter models. These results indicate that robust multiparameter EFWI depends not only on the optimization algorithm or neural architecture, but fundamentally on the compatibility among wave physics, acquisition geometry, physical parameterization, and model representation. By combining an INR-constrained model space with a physically favorable elastic parameterization, the proposed framework provides a practical route toward more robust multiparameter elastic inversion of complex synthetic and field seismic data.

\section{METHODOLOGY}
\subsection{Multiparameter elastic full-waveform inversion}
We consider 2D isotropic elastic wave propagation in the velocity–stress formulation, and the wavefield can be expressed via the following expression
\begin{equation}
\label{eq:elasticwave1}
\rho \frac{\partial \mathbf{v}}{\partial t}=\nabla\cdot\boldsymbol{\sigma}+\mathbf{f},
\end{equation}
\begin{equation}
\label{eq:elasticwave2}
\frac{\partial \boldsymbol{\sigma}}{\partial t}=
\lambda(\nabla\cdot\mathbf{v})\mathbf{I}
+\mu\left[\nabla\mathbf{v}+(\nabla\mathbf{v})^{T}\right],
\end{equation}
where \(\mathbf{v}\) is particle velocity, \(\boldsymbol{\sigma}\) is the stress tensor, \(\rho\) is density, \(\lambda\) and \(\mu\) are the Lamé parameters, and \(\mathbf{f}\) denotes the seismic source. The conventional elastic parameters are related through
\begin{equation}
\label{eq:lame1}
\lambda=\rho(V_P^2-2V_S^2), \qquad
\mu=\rho V_S^2.
\end{equation}
Given observed multicomponent data $\mathbf{d}^{\rm obs}$, elastic FWI estimates a set of model parameters \(\mathbf{m}\) by minimizing the waveform discrepancy between the observed data and the calculated data $\mathbf{d}^{\rm cal}$,
\begin{equation}
\label{eq:objefwi}
\Phi(\mathbf{m})=\frac{1}{2}\sum_s\left\|\mathbf{W}\left[\mathbf{d}^{\rm cal}_s(\mathbf{m})-\mathbf{d}^{\rm obs}_s\right]\right\|_2^2 ,
\end{equation}
where \(s\) denotes the source index, \(\mathbf{W}\) represents optional data weighting, windowing, or preprocessing, and other misfit functions can be incorporated without changing the model representation developed below.

\par Numerically, the gradient with respect to each physical parameter can be efficiently evaluated using the adjoint-state method \citep{pratt1998}. For a model vector $\mathbf{m}=[m_1,m_2,\ldots,m_K]$, the conventional grid-based update takes the form
\begin{equation}
\label{eq:update}
\mathbf{m}^{k+1}=\mathbf{m}^{k}-\alpha_k\mathbf{P}_k\nabla_{\mathbf{m}}\Phi,
\end{equation}
where $\alpha_k$ is the step length and $\mathbf{P}_k$ denotes an optional gradient preconditioner or optimization operator. In multiparameter EFWI, however, the gradients of different parameters can exhibit substantially different amplitudes and overlapping sensitivity patterns. Consequently, a perturbation associated with one physical property may be partially explained by another, producing parameter crosstalk and degrading the reliability of the recovered elastic models.

\subsection{Physical parameterization of multiparameter EFWI}
\begin{figure*}
\centering
\includegraphics[width=1.0\columnwidth]{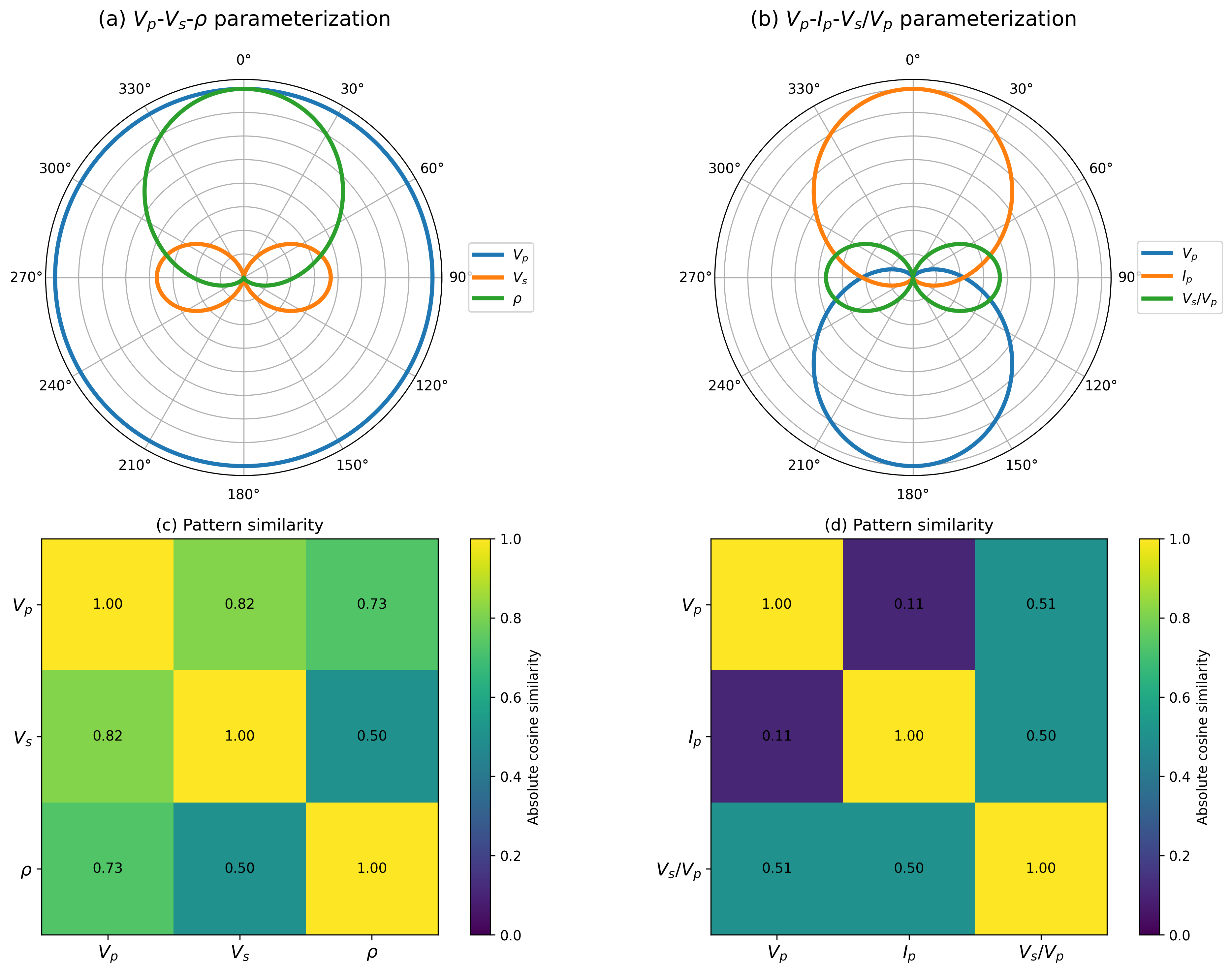}
\caption{Comparison of analytical P-wave radiation patterns and their pairwise similarities for two elastic parameterizations. (a) Radiation patterns for the $V_p-V_s-\rho$ parameterization. (b) Radiation patterns for the $V_p-I_p-V_s/V_p$ parameterization. (c) Absolute cosine-similarity matrix of the radiation patterns in (a). (d) Corresponding similarity matrix for the parameterization in (b). Lower off-diagonal similarity indicates improved parameter separability and reduced potential cross-talk.}
\label{fig:pattern}
\end{figure*}

An isotropic elastic medium can be represented in multiple ways; mathematical equivalence of these parameter sets does not imply equivalent behavior during inversion. Changing variables transforms the model gradient according to the chain rule. If $\mathbf{m}$ and $\mathbf{q}$ denote two parameterizations related through $\mathbf{m}=\mathbf{g}(\mathbf{q})$, then
\begin{equation}
\label{eq:chain}
\nabla_{\mathbf{q}}\Phi
=
\left(
\frac{\partial\mathbf{m}}
{\partial\mathbf{q}}
\right)^T
\nabla_{\mathbf{m}}\Phi.
\end{equation}
Reparameterization therefore changes the orientation and relative scaling of the gradients in model space, even though the underlying elastic wave equation remains unchanged. This directly affects parameter crosstalk and optimization conditioning. To determine a better parameterization, we analyze two main choices $(V_P,V_S,\rho)$ and $(V_P,I_P,Vs/Vp)$. In a simple case with $ V_P=2500$ m/s, $ V_S=1200$ m/s, and $\rho=1800$ mg/cc, we estimate the P-wave radiation patterns \citep{operto2013,panwy2018} and their corresponding similarities. Figure~\ref{fig:pattern} shows that, for the conventional $(V_P,V_S,\rho)$ parameterization, the radiation patterns overlap considerably, particularly between $V_P$ and $V_S$, whose absolute cosine similarity reaches 0.82. The similarity between Vp and $\rho$ is also relatively high (0.73), indicating strong coupling among the model parameters and a greater potential for crosstalk during multiparameter inversion. In contrast, the $(V_P,I_P,Vs/Vp)$ parameterization produces more distinct angular responses. In particular, the radiation patterns of $V_P$ and $I_P$ are nearly complementary, with a much lower cosine similarity of 0.11. The similarities between $Vs/Vp$ and the other two parameters are also reduced to approximately 0.50. Overall, the lower off-diagonal similarities indicate improved parameter separability for the $(V_P,I_P,Vs/Vp)$ parameterization, which is expected to reduce parameter trade-offs and mitigate crosstalk in EFWI. Therefore, in this study, we adopt the second parameterization $(V_P,I_P,R_{SP})$ for inversion, where $R_{SP}=Vs/Vp$. Then the quantities required by the elastic wave-equation solver are recovered through $V_S=R_{SP}V_P$ and $\rho=I_P/V_P$ followed by $\mu=\rho V_S^2$ and $\lambda=\rho(V_P^2-2V_S^2)$. This transformation is differentiable and can therefore be incorporated directly between the INR and elastic forward solver. The complete computational chain becomes
\begin{equation}
\label{eq:gradientchain}
(x,z)
\xrightarrow{\mathrm{INR}}
(V_P,I_P,V_S/V_P)
\xrightarrow{\mathrm{reparameterization}}
(V_P,V_S,\rho)
\xrightarrow{\mathrm{elastic\ solver}}
\mathbf{d}^{\rm cal}
\xrightarrow{\Phi}
\text{model update}.
\end{equation}
Consequently, both the neural representation and physical reparameterization participate in the gradient propagation and are optimized consistently within a single inversion framework.
The motivation for physical reparameterization is to identify variables whose waveform sensitivities are sufficiently complementary over the angular and wave-mode coverage provided by the acquisition. Radiation-pattern analysis provides a useful first-order diagnostic because the scattering patterns associated with individual parameters reveal the angles and wave modes through which those parameters can be constrained. Parameter pairs with strongly overlapping radiation patterns are intrinsically more difficult to separate, whereas complementary sensitivities can reduce parameter trade-offs. The $(V_P,I_P,V_S/V_P)$ formulation is motivated by separating three related aspects of the elastic model: compressional propagation, impedance-related contrasts, and relative shear-to-compressional behavior that can reveal lithological variations and thereby facilitate reservoir identification and characterization. More importantly, this parameterization exhibits improved parameter separability and reduced sensitivity overlap, making it less susceptible to parameter crosstalk during multiparameter inversion. Consequently, we are going to combine this parameterization with multiscale INR framework, where the parameter-specific branches further promote independent representations of different model properties. The combination of a better-conditioned parameterization and multiscale INR therefore provides a more effective way to suppress parameter crosstalk and improve the robustness of EFWI. More details regarding multiscale INR EFWI can be found in the next section.

Importantly, we do not assume that $(V_P,I_P,V_S/V_P)$ is a universally optimal parameterization. Parameter recoverability depends on the interaction among radiation patterns, scattering angles, acquisition aperture, component availability, wave modes, frequency content, background properties, and geological structure. For example, surface acquisition with limited angular coverage and multicomponent OBN acquisition may provide substantially different sensitivity to the same parameter set. The preferred parameterization should therefore be regarded as acquisition- and geology-dependent. This observation leads to the central methodological principle of this study: robust multiparameter EFWI requires consideration of two coupled parameterizations. The INR determines how the spatial model is represented and constrained, whereas the physical parameterization determines how elastic information is partitioned among the inverted variables. The INR can improve spatial coherence and stabilize the model space, but it cannot eliminate an ambiguity imposed by the elastic physics. Conversely, a favorable physical parameterization can reduce crosstalk but does not by itself prevent unstable spatial updates. We therefore jointly exploit implicit model representation and physically informed reparameterization to improve the robustness of multiparameter EFWI.

\subsection{Multiscale implicit neural representation for elastic FWI}
Conventional FWI treats the value at each model grid point as an independent optimization variable. We instead represent the elastic model using a coordinate-based implicit neural representation (INR). A neural function $f_{\boldsymbol{\theta}}$ maps spatial coordinates to the physical properties,
\begin{equation}
\label{eq:inr}
\mathbf{m}(\mathbf{x})=f_{\boldsymbol{\theta}}\left[\gamma(\mathbf{x})\right],
\end{equation}
where $\mathbf{x}=(x,z)$, $\gamma(\mathbf{x})$ denotes the coordinate representation or encoding, and $\boldsymbol{\theta}$ contains the trainable network parameters. For a three-parameter elastic inversion, $f_{\boldsymbol{\theta}}$ aims to represent different variables (e.g., Vp, Vs, and density) simultaneously via $f_{\boldsymbol{\theta}}:(x,z)\rightarrow(m_1,m_2,m_3)$. Then, based on the generated models, we can use a wave equation solver to predict the seismic response, and the forward processing can therefore be written as
\begin{equation}
\label{eq:forward}
\mathbf{d}^{\rm cal}=\mathcal{F}\left[f_{\boldsymbol{\theta}}(\gamma(\mathbf{x}))\right],
\end{equation}
and the inversion aims to optimize the network parameters so that the implicit neural representation can describe the model that best explains the observed data. Mathematically, this process can be expressed as
\begin{equation}
\label{eq:inrupdate}
\boldsymbol{\theta}^{*}=\underset{\boldsymbol{\theta}}{\arg\min}\;\Phi\left(\mathcal{F}[
f_{\boldsymbol{\theta}}(\gamma(\mathbf{x}))
],\mathbf{d}^{\rm obs}\right).
\end{equation}

Rather than directly updating millions of independent model samples, the waveform misfit is propagated through the elastic wave-equation solver and subsequently through the INR,
\begin{equation}
\label{eq:gradient}
\frac{\partial\Phi}{\partial\boldsymbol{\theta}}=\frac{\partial\Phi}{\partial\mathbf{m}}
\frac{\partial\mathbf{m}}{\partial\boldsymbol{\theta}}.
\end{equation}
The first factor contains the elastic-waveform sensitivity to the physical model, whereas the second maps that sensitivity into the neural parameter space. The resulting optimization therefore retains the physics of EFWI while changing the space in which the model is represented and updated. This distinction is important. INR does not replace the elastic forward physics, nor does it require a pretrained model or a training database. Instead, the neural network continuously parameterizes the subsurface model and is optimized specifically for the observed seismic data. Because neighboring model values are generated through a shared coordinate-dependent function, the representation introduces spatial coupling and an implicit model-space constraint. This property can suppress poorly constrained short-wavelength fluctuations and facilitate coherent multiscale reconstruction. However, it does not by itself remove the physical crosstalk inherent in multiparameter EFWI.
\par
\begin{figure*}
\centering
\includegraphics[width=1.0\columnwidth]{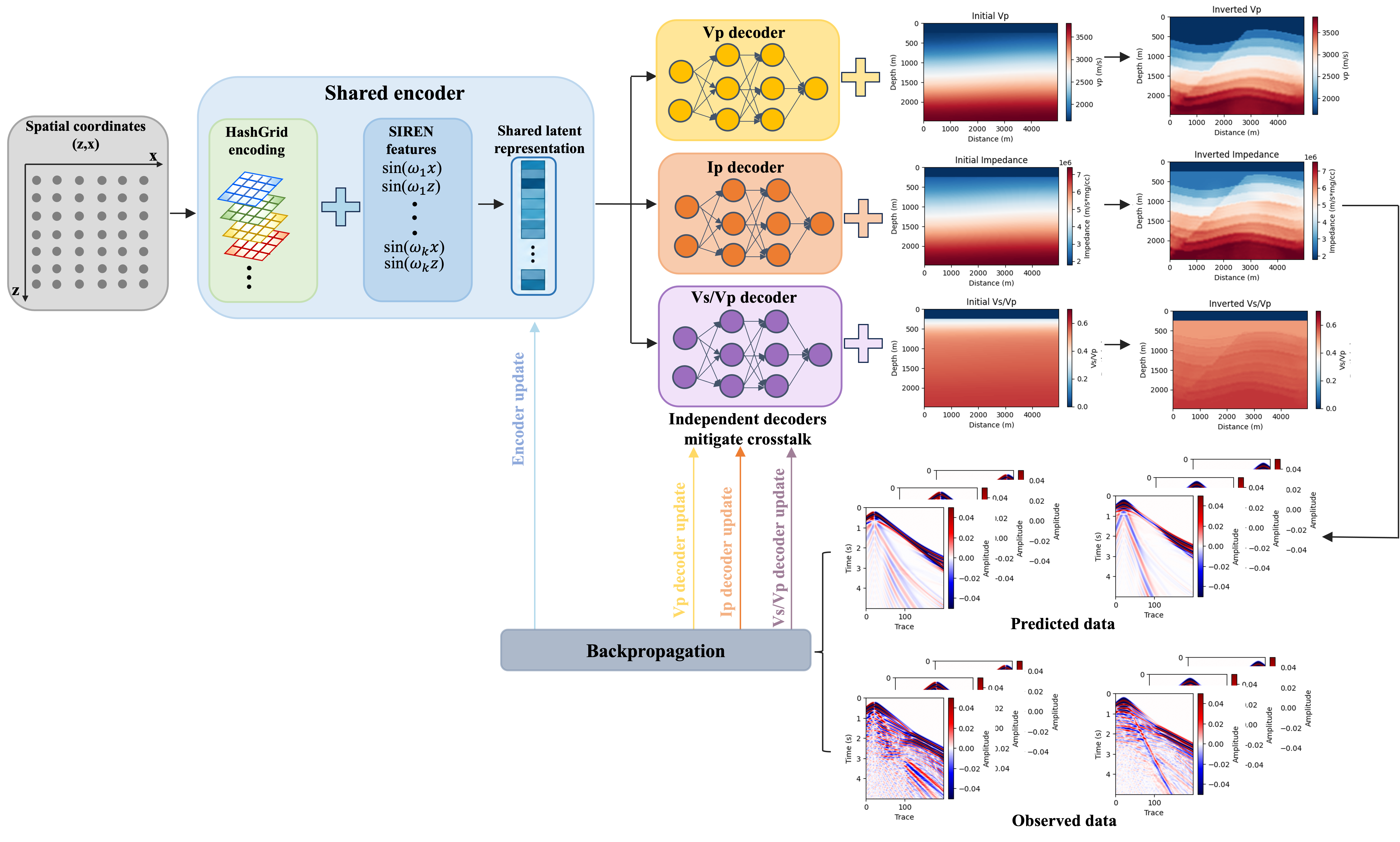}
\caption{A schematic illustration of INR-EFWI, in which three independent branches are designed for different variables to suppress parameter crosstalk. }
\label{fig:sketch}
\end{figure*}

 However, conventional INR-based FWI commonly reconstructs multiple subsurface properties from a shared latent representation, which may limit its ability to disentangle parameter-specific sensitivities and suppress crosstalk. To improve inversion stability, we develop a multiscale INR architecture in which several parameter-specific branches are connected to a shared encoder that extracts informative spatial features from the model coordinates as shown in Figure~\ref{fig:sketch}. The shared encoder combines multiresolution hash encoding \citep{wangsw2025} with SIREN-based sinusoidal representations \citep{sitzmann2020}. Specifically, the hash encoder contains 16 resolution levels with two features per level and a base resolution of 50. In parallel, the SIREN pathway contains three sinusoidal layers with widths of 256, 256, and 128, respectively, and uses $\omega_0=30$ for all layers to model continuous spatial variations. By integrating these complementary representations, the encoder can simultaneously characterize smooth background trends and fine-scale geological structures, while the separate output branches allow different elastic parameters to be reconstructed through dedicated mappings, thereby helping reduce crosstalk during inversion. Mathematically, let $\mathbf{h}_{\mathrm{hash}}(\mathbf{x})$ and $\mathbf{h}_{\mathrm{siren}}(\mathbf{x})$ denote the features extracted by the multiresolution hash encoder and the SIREN pathway, respectively. The two feature vectors are concatenated to form the shared latent representation $\mathbf{h}(\mathbf{x})$
\begin{equation}
\label{eq:hash}
\mathbf{h}(\mathbf{x})=[\mathbf{h}_{\mathrm{hash}}(\mathbf{x});\mathbf{h}_{\mathrm{siren}}(\mathbf{x})].
\end{equation}
Subsequently, the shared vector will go through independent decoder branches with the same architecture. In detail, each decoder consists of two hidden layers with 128 and 64 neurons, followed by a single output linear layer, with GELU activations applied between hidden layers. Although the three decoders receive the same shared representation, their trainable parameters are independent, allowing each branch to learn parameter-specific mappings. This design provides a common structural representation of the subsurface while preserving sufficient flexibility for individual elastic parameters, which helps  mitigate crosstalk during multiparameter inversion. For each property, the updated model can be written as 
\begin{equation}
\label{eq:updatevp}
v_p(\mathbf{x})=v_{p0}(\mathbf{x})+s_{v_{p}}\mathbf{D}_{v_{p}}\left(\mathbf{h}(\mathbf{x})\right),
\end{equation}
\begin{equation}
\label{eq:updateip}
I_p(\mathbf{x})=I_{p0}(\mathbf{x})+s_{I_{p}}\mathbf{D}_{I_{p}}\left(\mathbf{h}(\mathbf{x})\right),
\end{equation}
\begin{equation}
\label{eq:updateratio}
R_{SP}(\mathbf{x})=R_{SP0}(\mathbf{x})+s_{R_{SP}}\mathbf{D}_{R_{SP}}\left(\mathbf{h}(\mathbf{x})\right),
\end{equation}
where $\mathbf{D}_{v_{p}}$, $\mathbf{D}_{I_{p}}$, and $\mathbf{D}_{R_{SP}}$ denote the decoder corresponding to different properties. $s_{v_{p}}$, $s_{I_{p}}$, and $s_{R_{SP}}$ represent the scaling factors used to restore the physical magnitudes of the corresponding model parameters. $v_{p0}(\mathbf{x})$, $I_{p0}(\mathbf{x})$, and $R_{SP0}(\mathbf{x})$ are the initial models used for inversion. Since different physical parameters are represented by separate decoder branches, the gradient associated with each parameter is backpropagated through its own parameter-specific network. Consequently, the updates of $V_p$, $I_p$, and $R_{SP}$ are controlled by independent sets of branch parameters, which reduces direct gradient interference among different model variables. This branch-wise optimization provides each model parameter with greater flexibility to adapt to its own sensitivity and spatial characteristics. As illustrated in Figure~\ref{fig:sketch}, the parameter-specific branches are independently refined by minimizing the misfit between the predicted and observed data, thereby reducing direct interference among different model updates.

\section{EXAMPLES}
\subsection{Marmousi example}
\begin{figure*}
\centering
\includegraphics[width=1.0\columnwidth]{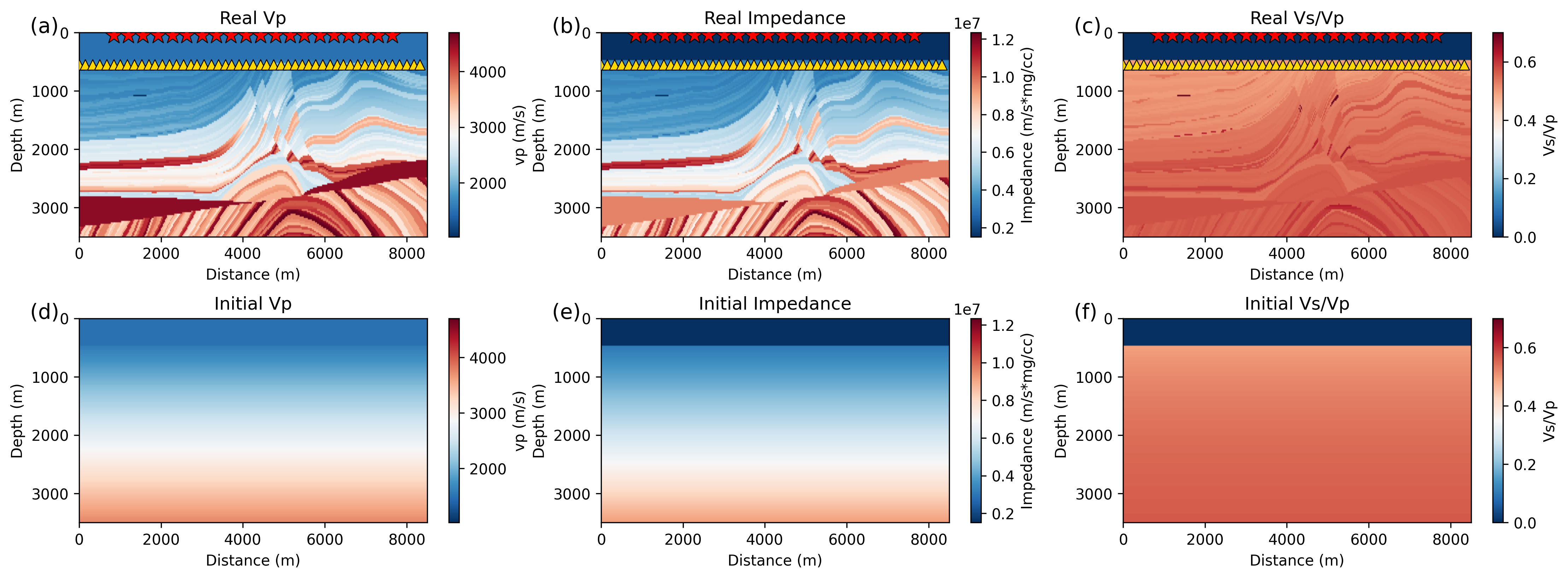}
\caption{(a)-(c) show the real Marmousi model, and (d)-(f) show initial Marmousi model used for inversion, including Vp, impedance, and Vs/Vp.}
\label{fig:marmini}
\end{figure*}

To evaluate the performance of the proposed INR-EFWI framework, Figures~\ref{fig:marmini}(a)-\ref{fig:marmini}(c) present the true models used for inversion. The model has $141 \times 341$ grid points with a spatial interval of 25 m. We uniformly distribute 20 shots along the surface and generate them using a 5 Hz Ricker wavelet. Each shot is recorded by 200 receivers. The acquisition geometry is overlaid on the velocity model, with red stars denoting the source locations and yellow triangles denoting the receiver locations. Moreover, Figures~\ref{fig:marmini}(d)-\ref{fig:marmini}(f) present the linear initial models of Vp, impedance, and Vs/Vp, respectively. These deliberately simplified initial models provide limited structural information, making the inversion more challenging and increasing the risk of convergence to local minima. We then conduct both conventional EFWI within an automatic-differentiation framework (AD-EFWI) and INR-EFWI using the same poor initial models. Specifically, we run both experiments for 500 iterations. Because Vp, impedance, and Vs/Vp have substantially different physical magnitudes and units, AD-EFWI uses parameter-specific learning rates to balance update scales across parameters. We set the learning rates for Vp, impedance, and Vs/Vp to 25, $1\times10^{4}$, and $1\times10^{-3}$, respectively. For INR-EFWI, the learning rates for the three parameter branches are all set to $5\times10^{-5}$ and $1\times10^{-4}$ for the encoder branch.

\par
\begin{figure*}
\centering
\includegraphics[width=1.0\columnwidth]{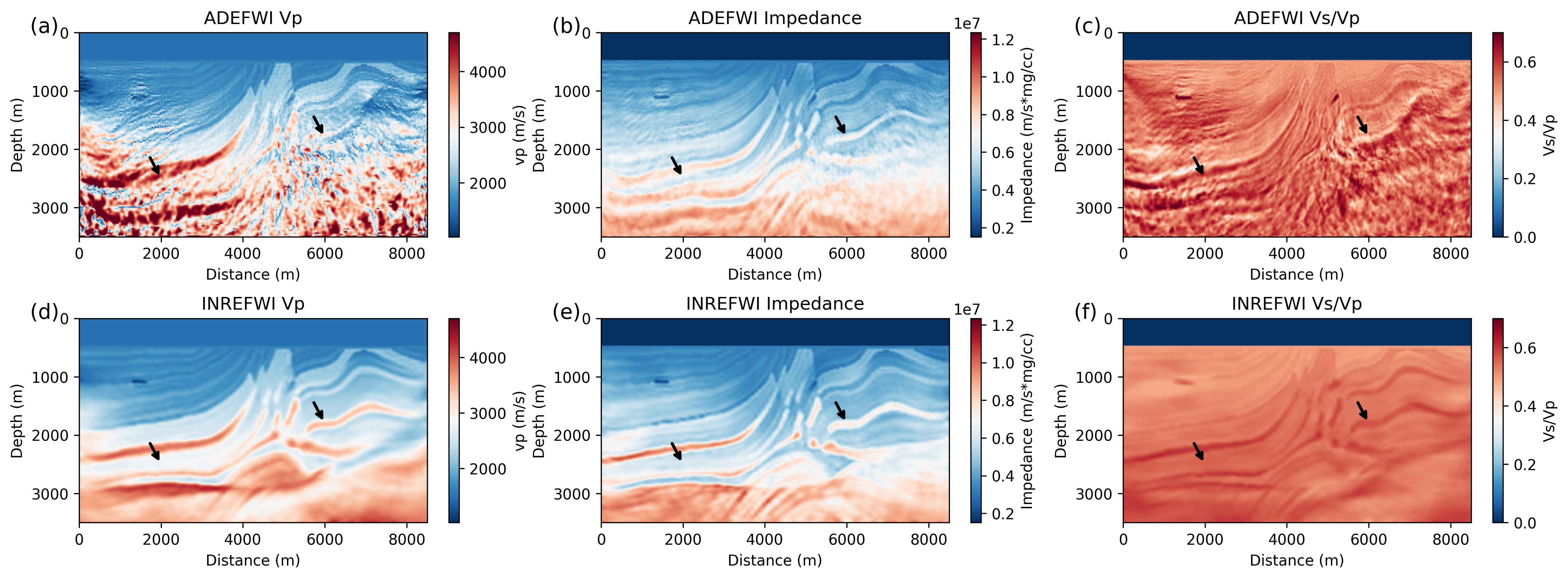}
\caption{(a)-(c) show the inverted Marmousi Vp, impedance, and Vs/Vp using AD-EFWI. (d)-(f) show the inverted Marmousi Vp, impedance, and Vs/Vp using INR-EFWI. }
\label{fig:marminv}
\end{figure*}

 Figure~\ref{fig:marminv} compares the inverted results obtained using conventional AD-EFWI and INR-EFWI. The AD-EFWI results in Figures~\ref{fig:marminv}(a)-\ref{fig:marminv}(c) contain pronounced artifacts and oscillatory features, especially in the deeper part of the model and near the complex structures highlighted by the arrows, indicating strong parameter cross-talk. As a result, several interfaces become blurred or discontinuous. By comparison, the INR-EFWI results shown in Figures~\ref{fig:marminv}(d)-\ref{fig:marminv}(f) show substantially reduced artifacts, improved structural continuity, and more coherent recovery of the major geological features, suggesting that the implicit neural representation helps suppress parameter cross-talk during multiparameter inversion. Additionally, for a quantitative comparison, we employ the mean absolute error (MAE), model signal-to-noise ratio (MSNR), and structural similarity index measure (SSIM) to further evaluate the inversion results. Table~\ref{table1} summarizes the quantitative results, with superior results highlighted in light blue. Compared with conventional AD-EFWI, INR-EFWI generally achieves lower MAE and higher MSNR and SSIM values, demonstrating improved model fidelity and better preservation of structural features.

\par
\begin{figure*}
\centering
\includegraphics[width=1.0\columnwidth]{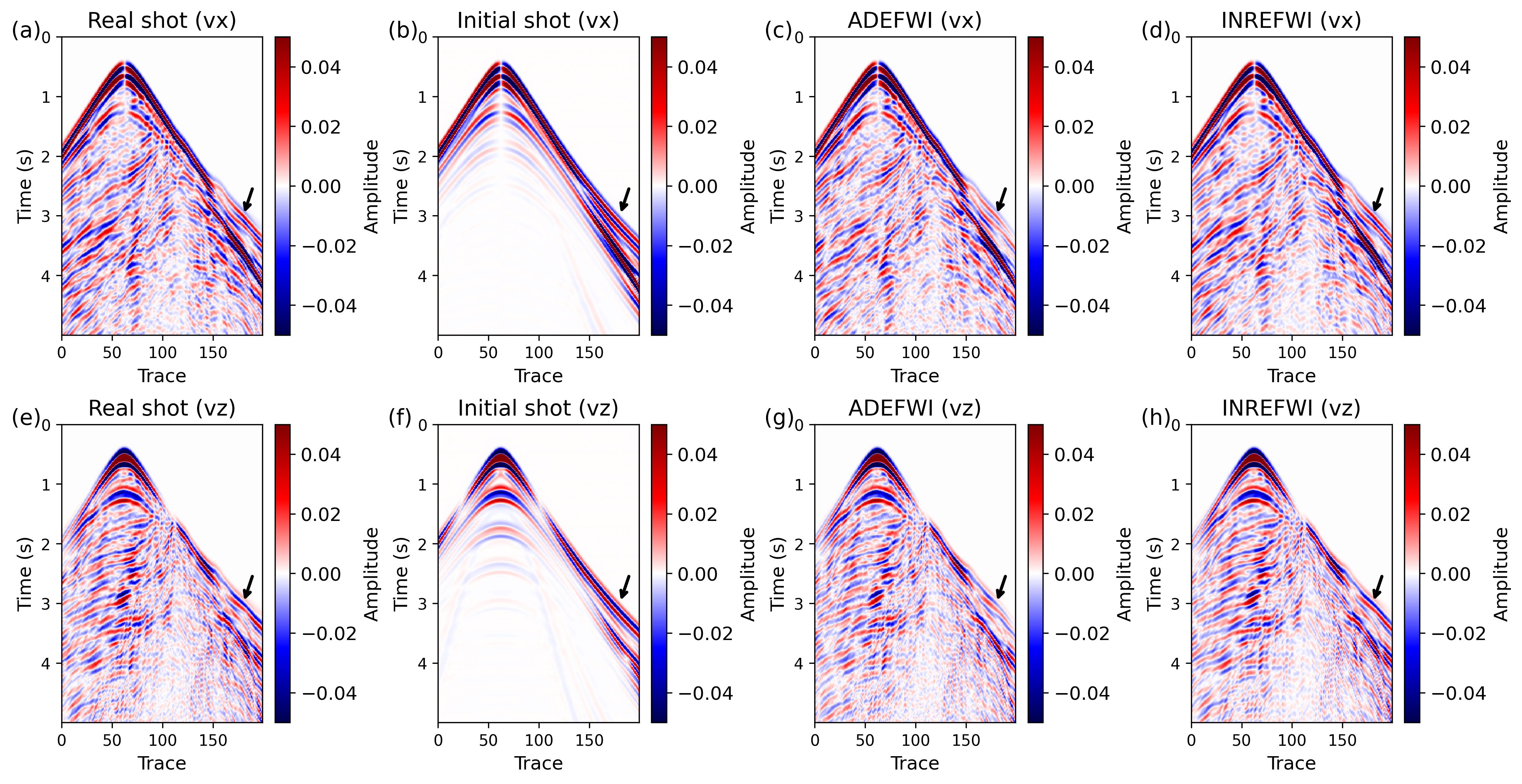}
\caption{Shot gathers of the $v_x$ component generated using (a) the true model, (b) the initial model, (c) the model inverted by AD-EFWI, and (d) the model inverted by INR-EFWI; and shot gathers of the $v_z$ component generated using (e) the true model, (f) the initial model, (g) the model inverted by AD-EFWI, and (h) the model inverted by INR-EFWI. }
\label{fig:marmshots}
\end{figure*}

\begin{figure*}
\centering
\includegraphics[width=1.0\columnwidth]{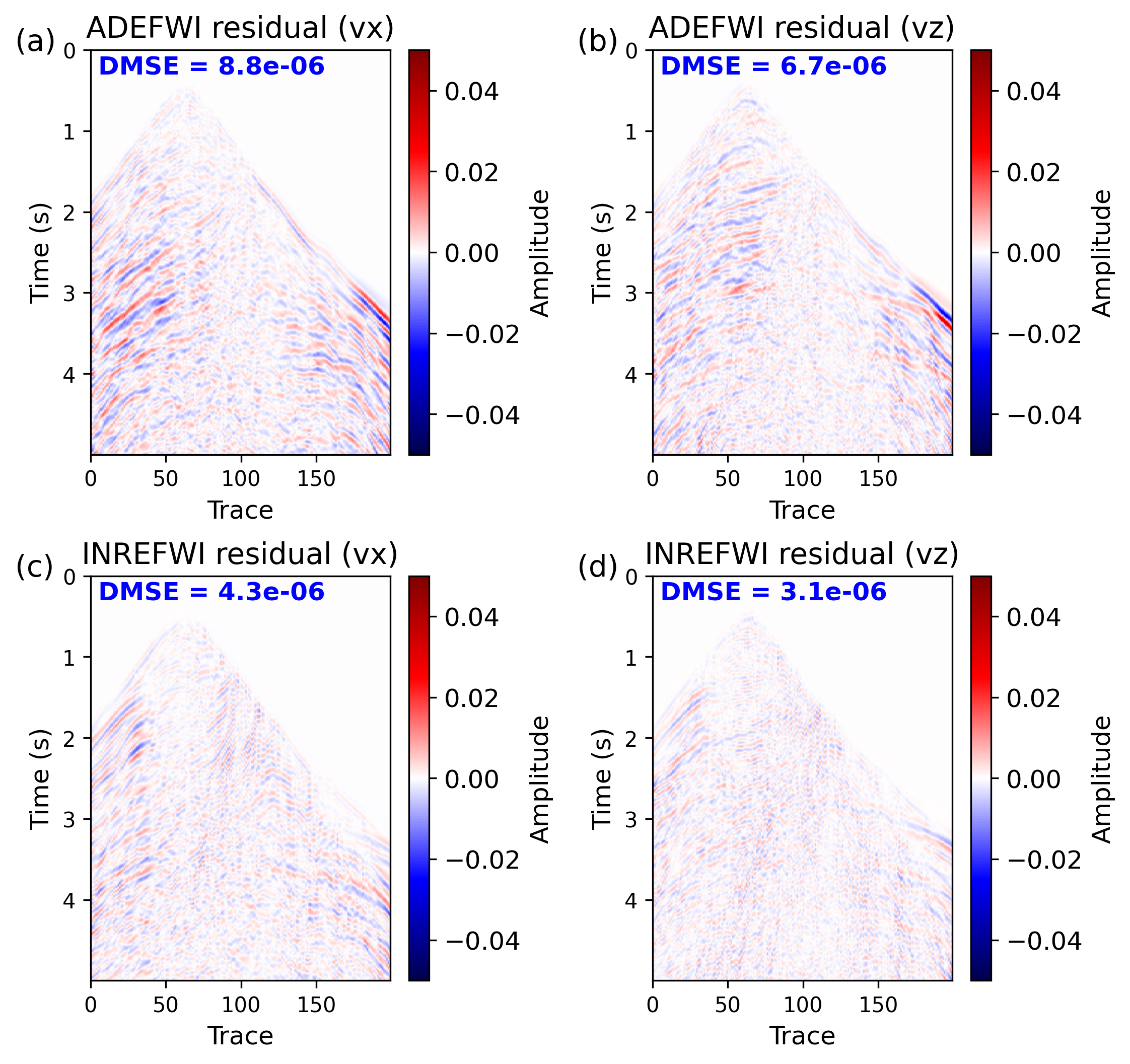}
\caption{Shot-gather residuals between the observed and predicted data for the $v_x$ and $v_z$ components. (a) and (b) Residuals obtained using the AD-EFWI-inverted model for the $v_x$ and $v_z$ components, respectively. (c) and (d) Corresponding residuals obtained using the INR-EFWI-inverted model. The DMSE values are shown in each panel. }
\label{fig:marmshotres}
\end{figure*}

 Moreover, beyond the comparison of the inverted models, we further assess the inversion performance in the data domain by comparing the shot gathers before and after inversion. Figure~\ref{fig:marmshots} shows the shot gathers of the $V_x$ and $V_z$ components, respectively. Due to the strong smoothness of the initial model, most reflected events are missing in Figures~\ref{fig:marmshots}(b) and~\ref{fig:marmshots}(f). Figures~\ref{fig:marmshots}(c),~\ref{fig:marmshots}(d),~\ref{fig:marmshots}(g), and~\ref{fig:marmshots}(h) present the shot gathers generated using the AD-EFWI and INR-EFWI inverted models, respectively. Compared with the real shot gathers, the diving-wave amplitudes produced by AD-EFWI exhibit noticeable mismatches, suggesting that the recovered low-wavenumber background velocity is still inaccurate. Furthermore, Figure~\ref{fig:marmshotres} presents the data residuals of the $V_x$ and $V_z$ components between the ground-truth shot gathers and those generated using the inverted models. We also employ the data mean squared error (DMSE) to quantitatively evaluate the data-fitting accuracy. Compared with conventional AD-EFWI, INR-EFWI yields smaller residuals and lower DMSE values, indicating that the models recovered by INR-EFWI not only achieve higher model-space accuracy but also provide a better explanation of the observed seismic data.
\begin{figure*}
\centering
\includegraphics[width=1.0\columnwidth]{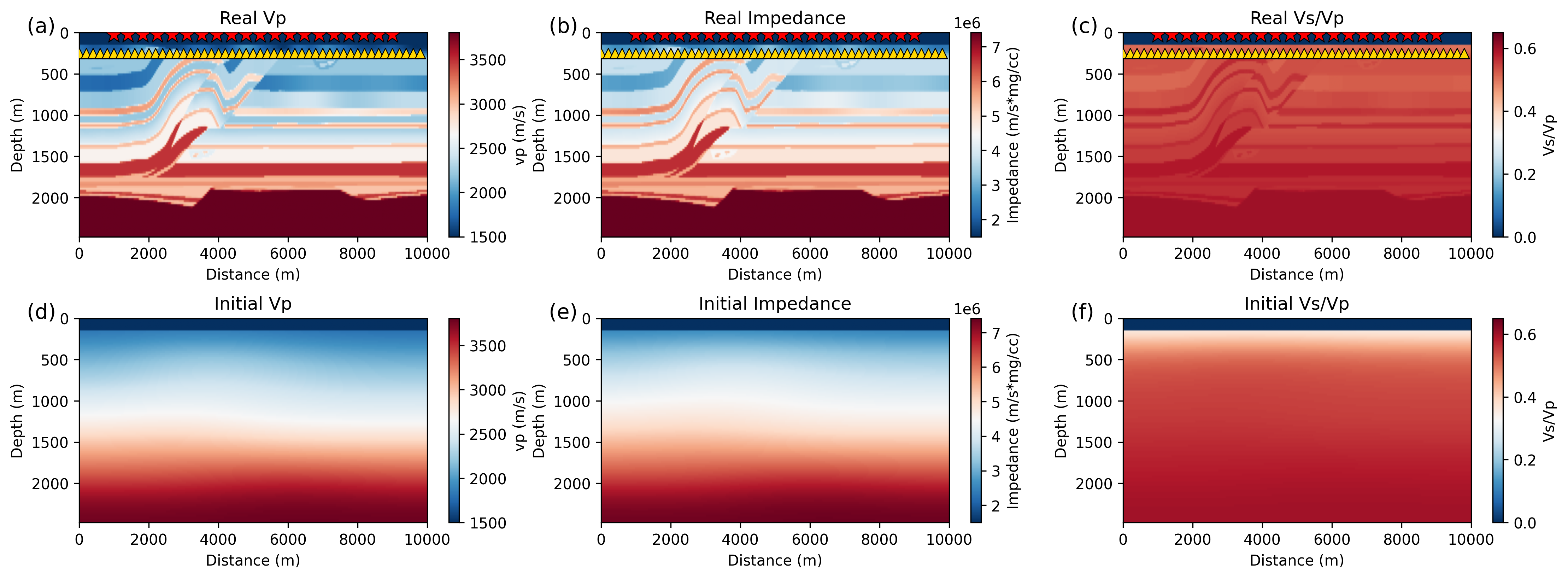}
\caption{(a)-(c) show the real Overthrust model, and (d)-(f) show the initial Overthrust model used for inversion, including Vp, impedance, and Vs/Vp.}
\label{fig:overini}
\end{figure*}
\begin{figure*}
\centering
\includegraphics[width=1.0\columnwidth]{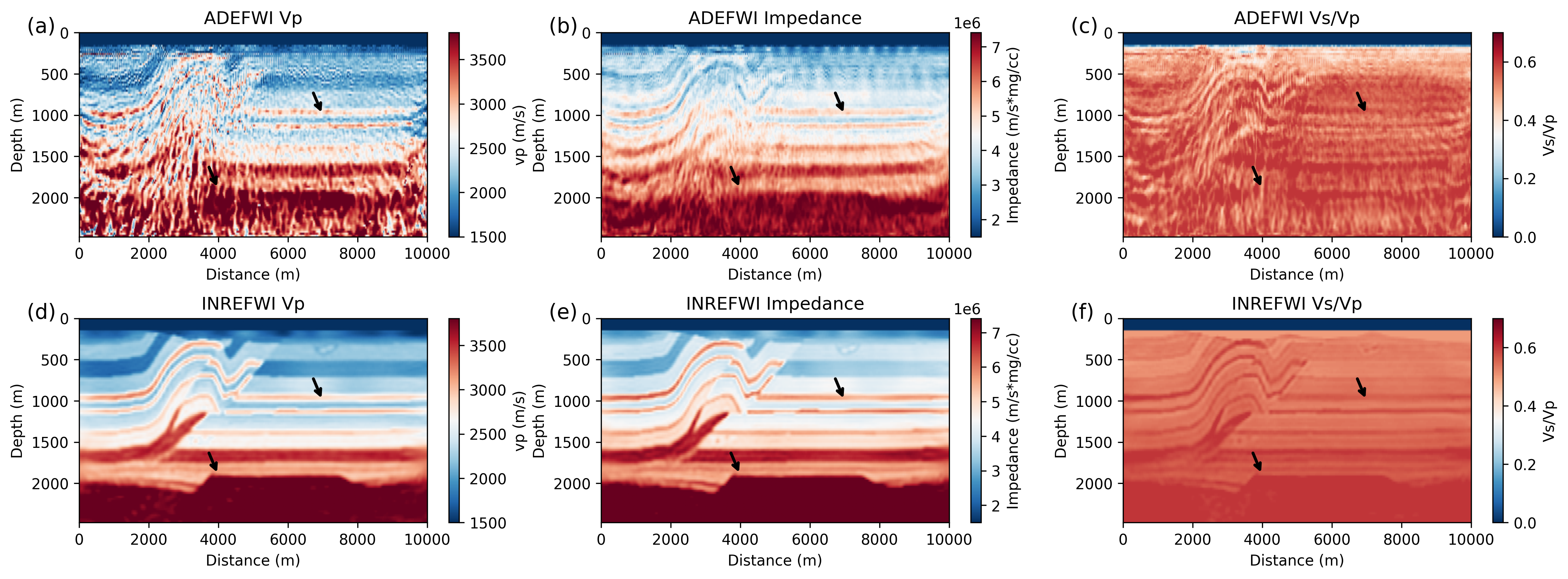}
\caption{(a)-(c) show the inverted Overthrust Vp, impedance, and Vs/Vp using AD-EFWI. (d)-(f) show the inverted Overthrust Vp, impedance, and Vs/Vp using INR-EFWI. }
\label{fig:overinv}
\end{figure*}
\subsection{Overthrust example}
\par Next, we use another widely used Overthrust model to further evaluate the effectiveness of the proposed method. Figure~\ref{fig:overini} shows the true and initial Overthrust models used for inversion. The model size is \(100 \times 401\) with a spatial interval of 25 m. As in the Marmousi experiment, we adopt the same acquisition geometry, source wavelet, and inversion parameter settings, including the learning rates for different model parameters. The source and receiver locations are overlaid on the model in Figure~\ref{fig:overini}, making the acquisition configuration directly visible. Figure~\ref{fig:overinv} shows the inverted results after 300 iterations using classic AD-EFWI and INR-EFWI for different variables (e.g., Vp, impedance, and Vs/Vp), respectively. In detail, AD-EFWI results show noticeable artifacts and parameter cross-talk, which blur interfaces and weaken structural continuity. In contrast, the INR-EFWI results shown in Figures~\ref{fig:overinv}(d)-\ref{fig:overinv}(f) recover the major geological structures more clearly, with sharper interfaces and better continuity. In particular, INR-EFWI reconstructs the structures highlighted by the arrows more accurately, indicating that it effectively suppresses parameter cross-talk and produces higher-quality inversion results than conventional AD-EFWI. In addition to the visual improvements, Table~\ref{table1} summarizes the MAE, MSNR, and SSIM values for the Overthrust model. Compared with AD-EFWI, INR-EFWI achieves lower MAE and higher MSNR and SSIM, further demonstrating improved reconstruction accuracy and structural fidelity.
\begin{figure*}
\centering
\includegraphics[width=1.0\columnwidth]{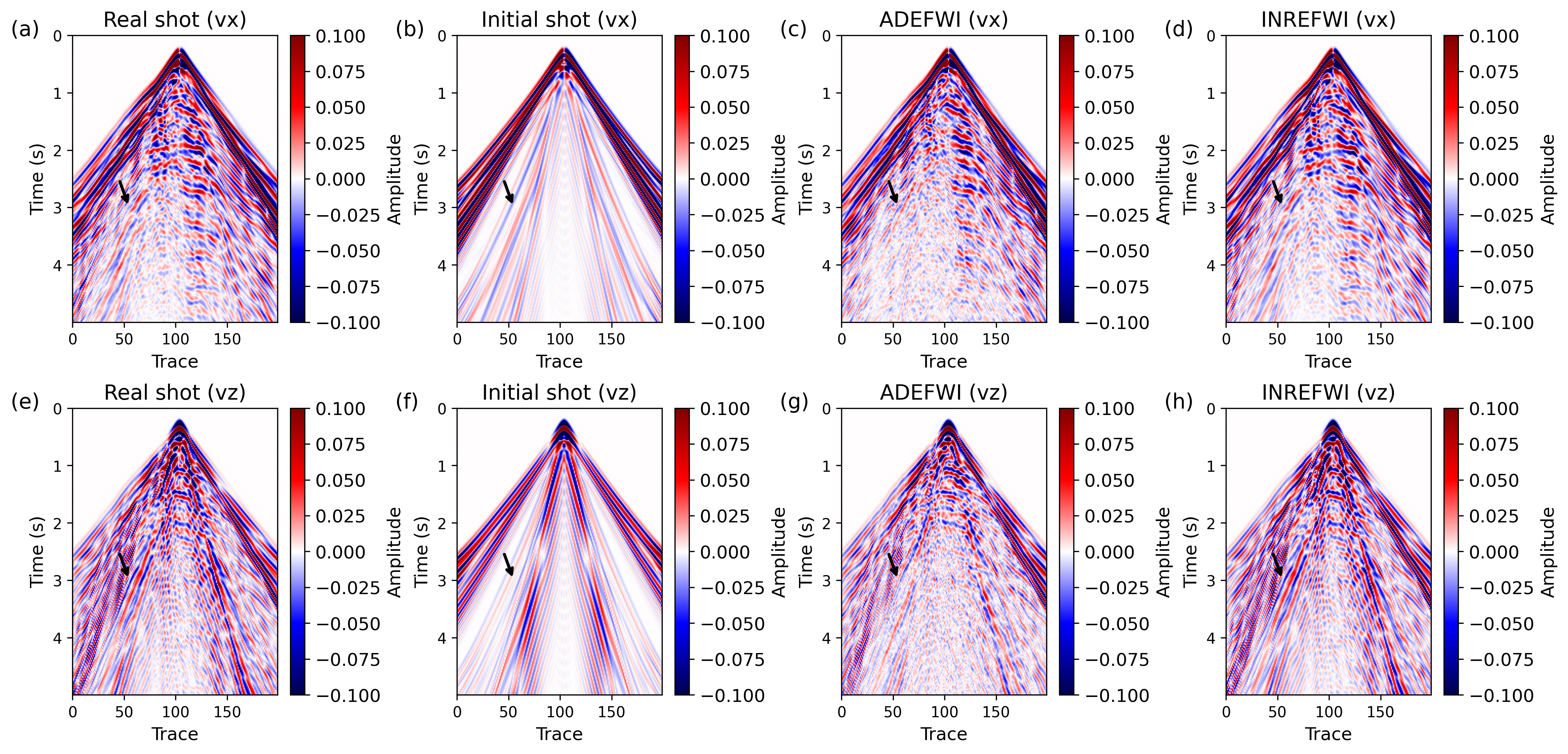}
\caption{Shot gathers of the $v_x$ component generated using (a) the true model, (b) the initial model, (c) the model inverted by AD-EFWI, and (d) the model inverted by INR-EFWI; and shot gathers of the $v_z$ component generated using (e) the true model, (f) the initial model, (g) the model inverted by AD-EFWI, and (h) the model inverted by INR-EFWI. }
\label{fig:overshot}
\end{figure*}
\begin{figure*}
\centering
\includegraphics[width=1.0\columnwidth]{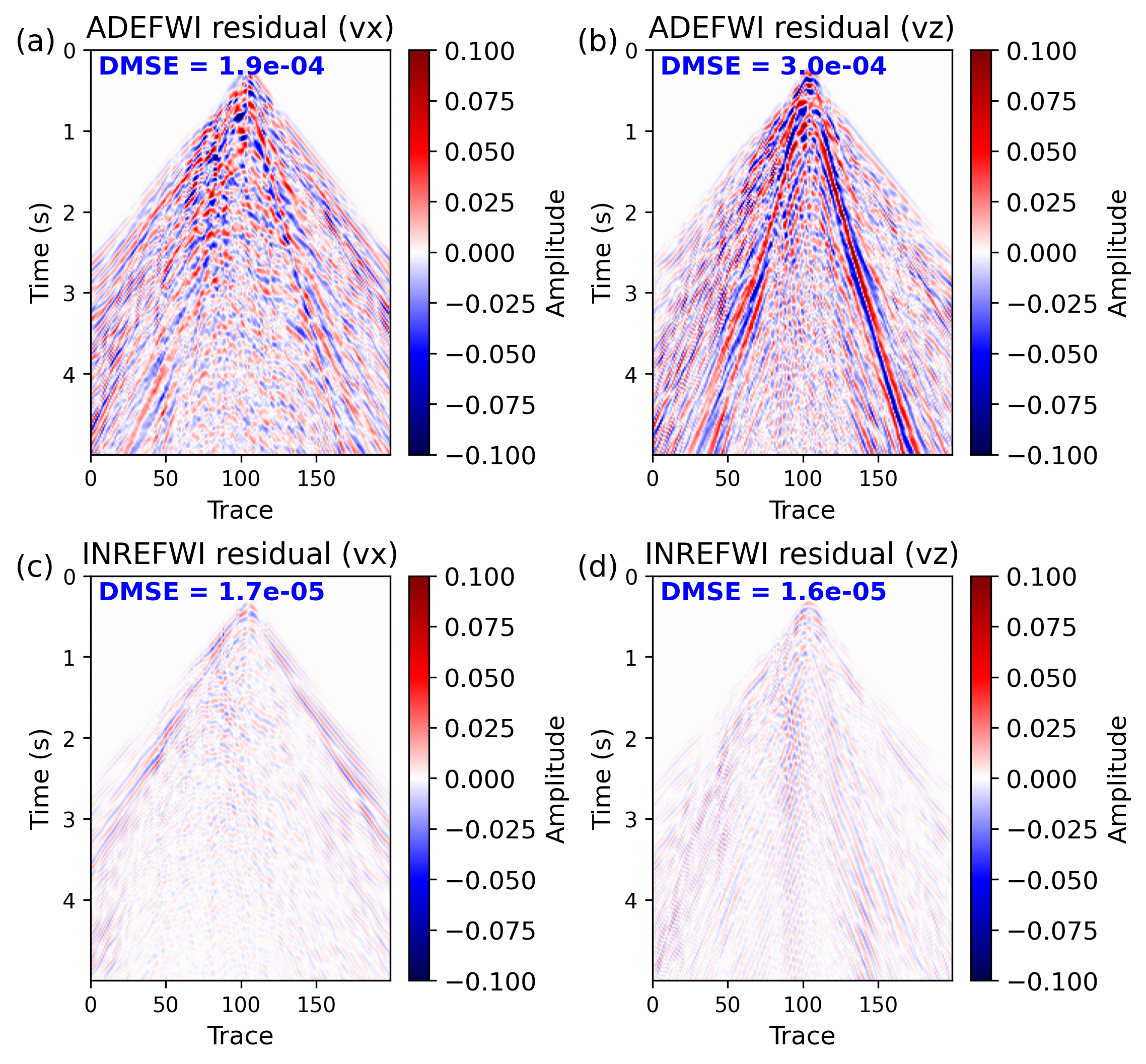}
\caption{Shot-gather residuals between the observed and predicted data for the $v_x$ and $v_z$ components. (a) and (b) Residuals obtained using the AD-EFWI-inverted model for the $v_x$ and $v_z$ components, respectively. (c) and (d) Corresponding residuals obtained using the INR-EFWI-inverted model. The DMSE values are shown in each panel. }
\label{fig:overshotres}
\end{figure*}

\par Figure~\ref{fig:overshot} compares the $v_x$ and $v_z$ components of the shot gathers before and after inversion for the Overthrust model. Compared with the shot gathers generated from the AD-EFWI results, those generated from the INR-EFWI results exhibit better agreement with the observed data, with improved event continuity and reduced amplitude mismatch in the regions indicated by the black arrows. Furthermore, Figure~\ref{fig:overshotres} compares the data residuals of the $v_x$ and $v_z$ components obtained using the AD-EFWI and INR-EFWI inverted models. As shown in Figures~\ref{fig:overshotres}(a) and~\ref{fig:overshotres}(b), the AD-EFWI results exhibit strong coherent residual energy over a wide range of traces and arrival times, indicating noticeable mismatches between the predicted and real data. In contrast, the residuals obtained using INR-EFWI in Figures~\ref{fig:overshotres}(c) and~\ref{fig:overshotres}(d) are significantly weaker and less coherent. This improvement is also reflected by the DMSE values: the $v_x$ residual decreases from 1.9 $\times10^{-4}$ to $1.7\times10^{-5}$, while the $v_z$ residual decreases from $3.0\times10^{-4}$ to $1.6\times10^{-5}$. These results demonstrate that INR-EFWI provides substantially improved data consistency and better explains the observed seismic waveforms.
\begin{table*}[!t]%
\centering %
\caption{Quantitative comparison of AD-EFWI and INR-EFWI for the Marmousi and Overthrust models.
The symbols $\uparrow$ and $\downarrow$ indicate that larger and smaller values are better, respectively.
\label{table1}}%
\begin{tabular*}{\textwidth}{@{\extracolsep\fill}llcccccc@{\extracolsep\fill}}
\toprule

& 
& \multicolumn{2}{c}{MAE $\downarrow$}
& \multicolumn{2}{c}{MSNR (dB) $\uparrow$}
& \multicolumn{2}{c}{SSIM $\uparrow$} \\

\cmidrule(lr){3-4}
\cmidrule(lr){5-6}
\cmidrule(lr){7-8}

Model
& Parameter
& AD-EFWI & INR-EFWI
& AD-EFWI & INR-EFWI
& AD-EFWI & INR-EFWI \\

\midrule

\multirow{3}{*}{Marmousi}

& $V_p$
& 412.02
& \cellcolor{blue!20}216.90
& 12.84
& \cellcolor{blue!20}15.71
& 0.32
& \cellcolor{blue!20}0.53 \\

& $I_p$
& 8.9 $\times 10^5$
& \cellcolor{blue!20}8.1$\times 10^5$
& 13.25
& \cellcolor{blue!20}14.15
& 0.35
& \cellcolor{blue!20}0.53\\

& $V_s/V_p$
& 0.033
& \cellcolor{blue!20}0.012
& 20.56
& \cellcolor{blue!20}27.82
& 0.99
& \cellcolor{blue!20}1 \\

\midrule

\multirow{3}{*}{Overthrust}

& $V_p$
& 231.53
& \cellcolor{blue!20}47.28
& 18.27
& \cellcolor{blue!20}31.07
& 0.38
& \cellcolor{blue!20}0.87 \\

& $I_p$
& 3.9$\times 10^5$
& \cellcolor{blue!20}1.4$\times 10^5$
& 19.79
& \cellcolor{blue!20}27.57
& 0.46
& \cellcolor{blue!20}0.90 \\

& $V_s/V_p$
& 0.036
& \cellcolor{blue!20}0.004
& 19.78
& \cellcolor{blue!20}39.28
& 0.99
& \cellcolor{blue!20}1 \\

\bottomrule
\end{tabular*}
\end{table*}

\subsection{Field example}
\begin{figure*}
\centering
\includegraphics[width=1.0\columnwidth]{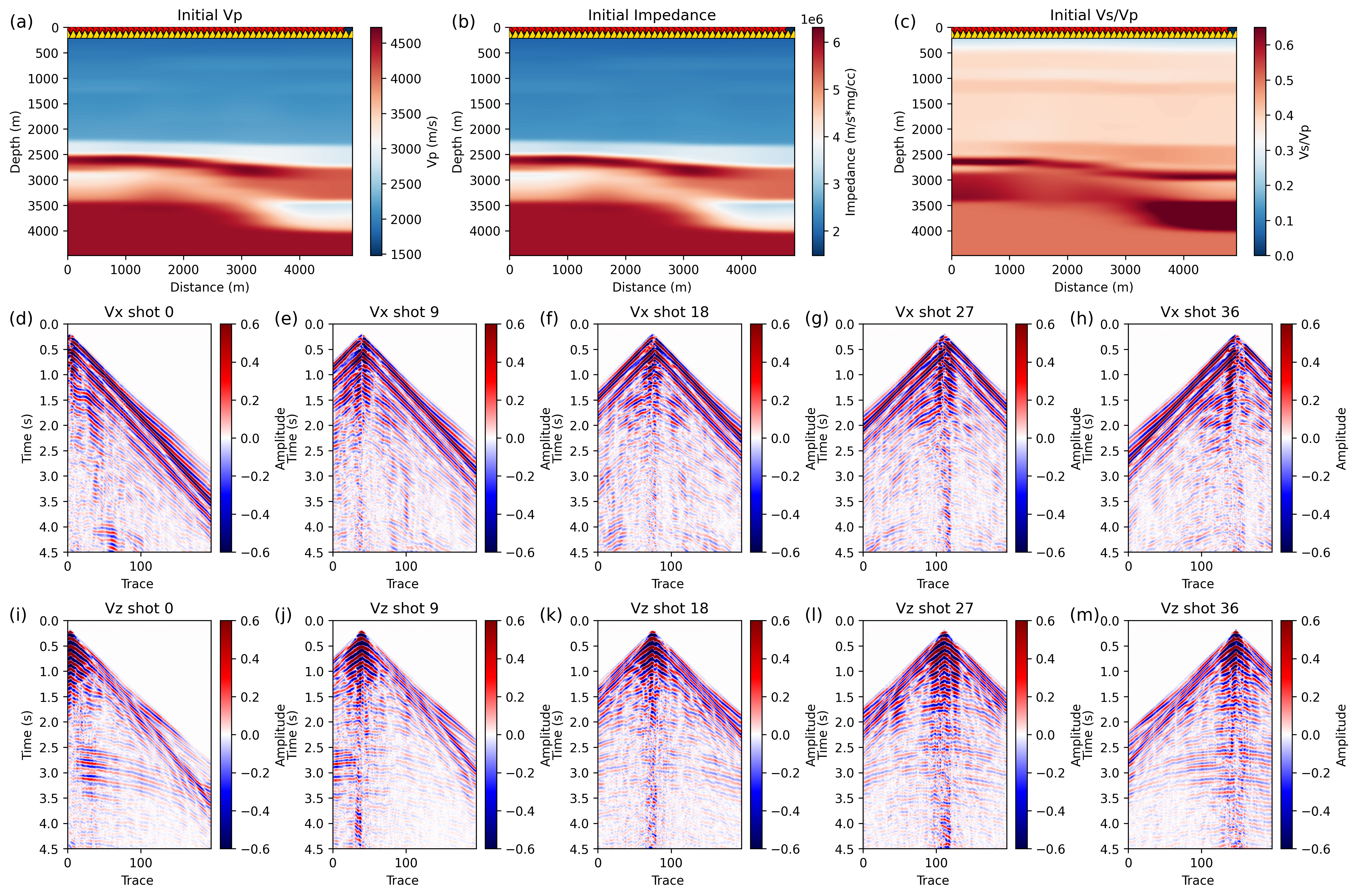}
\caption{(a)–(c) show the initial Vp, impedance, and Vs/Vp models used for the field-data inversion, respectively. (d)–(h) present representative $v_x$-component shot gathers from different source locations, and (i)–(m) present the corresponding $v_z$-component shot gathers.  }
\label{fig:fldini}
\end{figure*}

Finally, we apply the proposed method to an Ocean bottom cable (OBC) Volve dataset. The dataset contains both $v_x$ and $v_z$ components. In our experiment, we first filter the original dataset to extract signals ranging from 5 Hz to 12 Hz. Specifically, 48 shots with an interval of 100 m are used during inversion and each shot contains 197 traces with an interval of 25 m. Figure~\ref{fig:fldini} shows the initial models used for the field example with the acquisition geometry overlaid on the models. The models are discretized on a \(360 \times 394\) grid with a spatial interval of 12.5 m. The red stars denote the source locations, while the yellow triangles indicate the receiver locations. Figures~\ref{fig:fldini}(d)-\ref{fig:fldini}(h) show representative $v_x$-component shot gathers acquired at different source locations, and Figures~\ref{fig:fldini}(i)-\ref{fig:fldini}(m) show the corresponding $v_z$-component shot gathers. These shot gathers exhibit clear variations with source position and provide the observed data used for the subsequent inversion.
\begin{figure*}
\centering
\includegraphics[width=1.0\columnwidth]{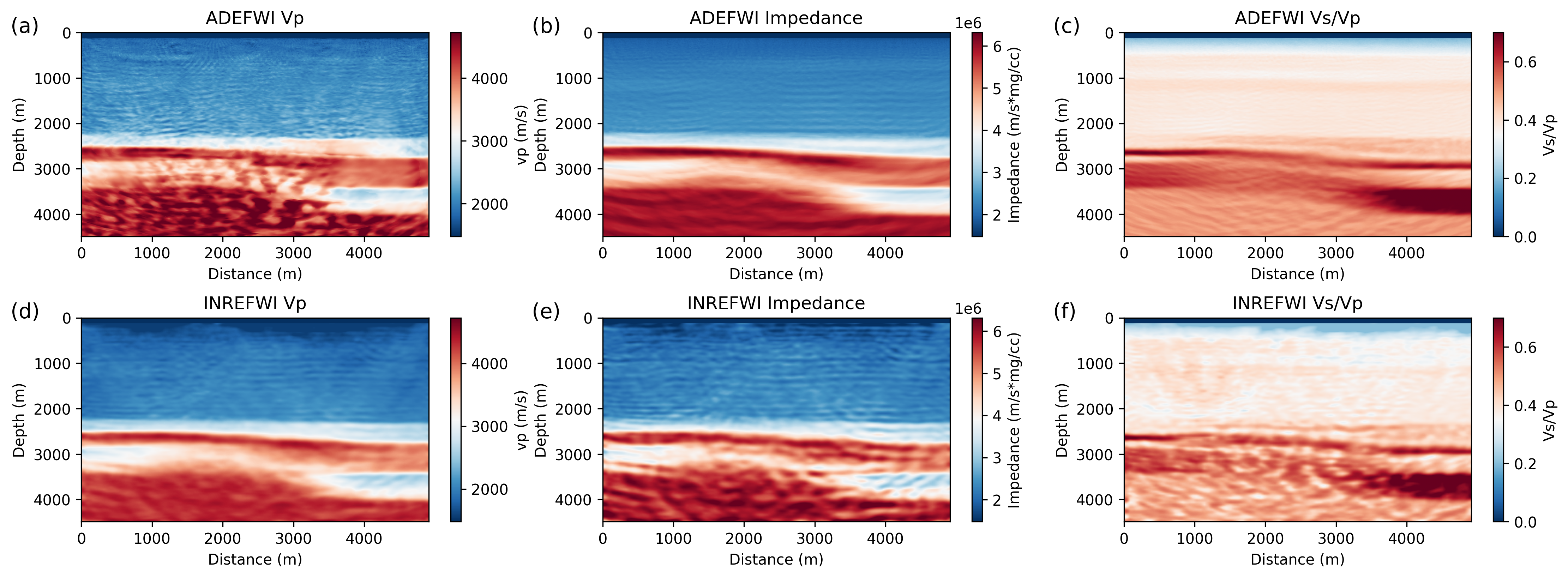}
\caption{(a)–(c) show the inverted Vp, impedance, and Vs/Vp models using AD-EFWI, respectively. (d)–(f) show the inverted Vp, impedance, and Vs/Vp models using INR-EFWI, respectively. }
\label{fig:fldinv}
\end{figure*}
\par
\begin{figure*}
\centering
\includegraphics[width=1.0\columnwidth]{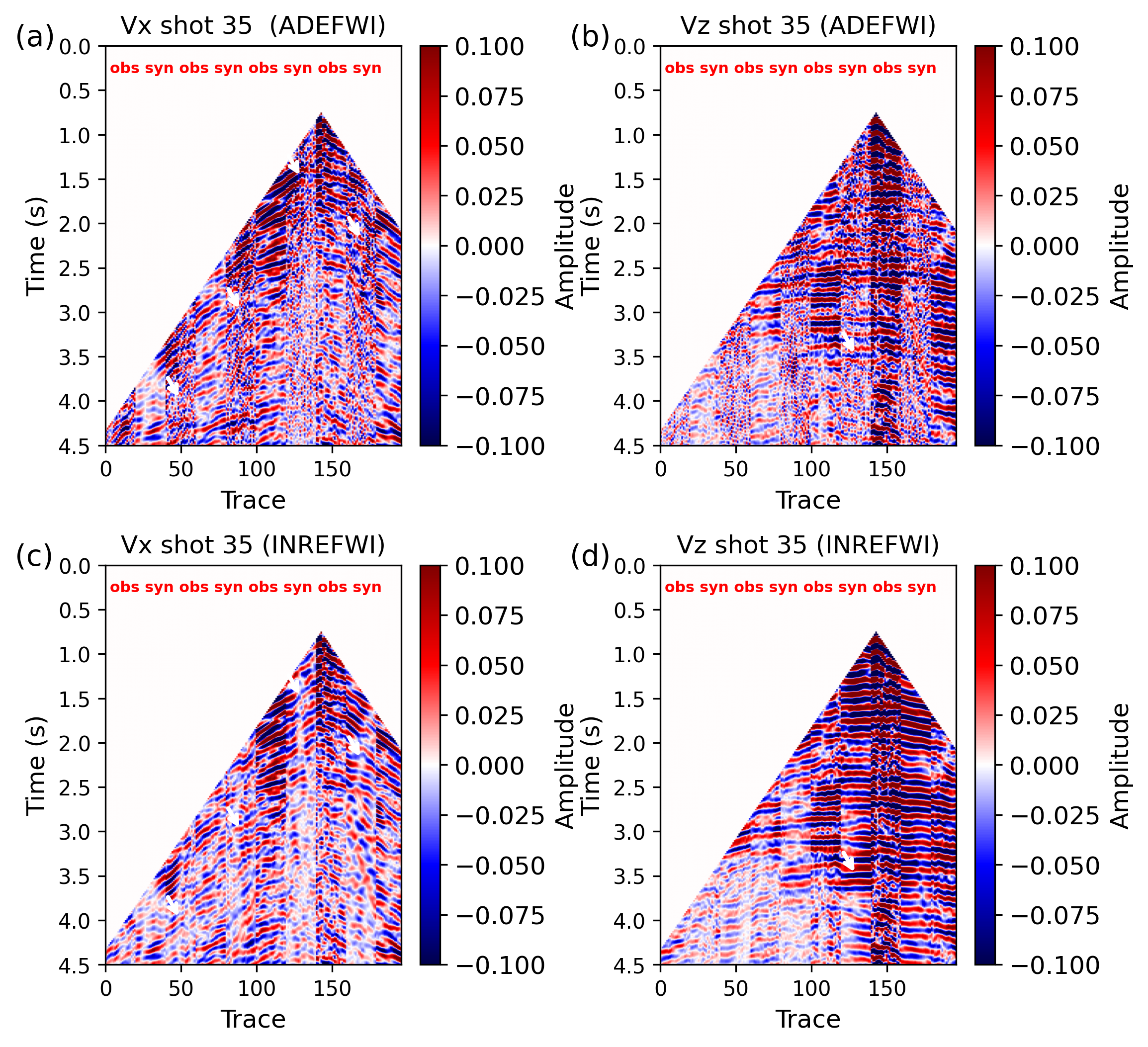}
\caption{(a) and (b) show the shot comparison ($v_x$ and $v_z$) corresponding to AD-EFWI. (c) and (d) show the shot comparison ($v_x$ and $v_z$) corresponding to INR-EFWI. }
\label{fig:fldshots}
\end{figure*}

 During inversion, we keep the learning rate unchanged and run 30 iterations for both AD-EFWI and INR-EFWI. Figure~\ref{fig:fldinv} presents the inverted results obtained using conventional AD-EFWI and INR-EFWI. Both methods enhance the model resolution relative to the initial models and recover several prominent reflectors that are not clearly visible before inversion. Both methods delineate shallow-to-intermediate structures well, while more obvious differences appear in the deeper portion of the model. The AD-EFWI results exhibit stronger oscillatory features and less coherent lateral variations, whereas the INR-EFWI results show more continuous interfaces and a smoother structural evolution with depth. The recovered Vp, impedance, and Vs/Vp models from INR-EFWI also show better structural consistency, suggesting reduced parameter cross-talk and improved geological plausibility. Because field-data applications lack reliable ground-truth models, we assess inversion performance through forward modeling. By comparing synthetic shot gathers generated from the inverted models with the observed data, we can directly evaluate the data consistency of the recovered models. Following this strategy, Figure~\ref{fig:fldshots} compares the $v_x$- and $v_z$-component shot gathers generated from the models inverted by AD-EFWI and INR-EFWI. For AD-EFWI, noticeable mismatches with the reference data appear in both components. In particular, the $v_x$ shot gather contains several fragmented, oscillatory events that are not well aligned with the reference wavefield, indicating spurious responses likely induced by parameter cross-talk. Similar discrepancies also appear in the $v_z$ component, although they are less pronounced. In contrast, the shot gathers generated from the INR-EFWI results show cleaner, more coherent events in both components, with better agreement with the reference data, suggesting reduced cross-talk-related artifacts and improved data fidelity.

\section{DISCUSSION}
\subsection{Sensitivity to low-frequency data}
\begin{figure*}
\centering
\includegraphics[width=1.0\columnwidth]{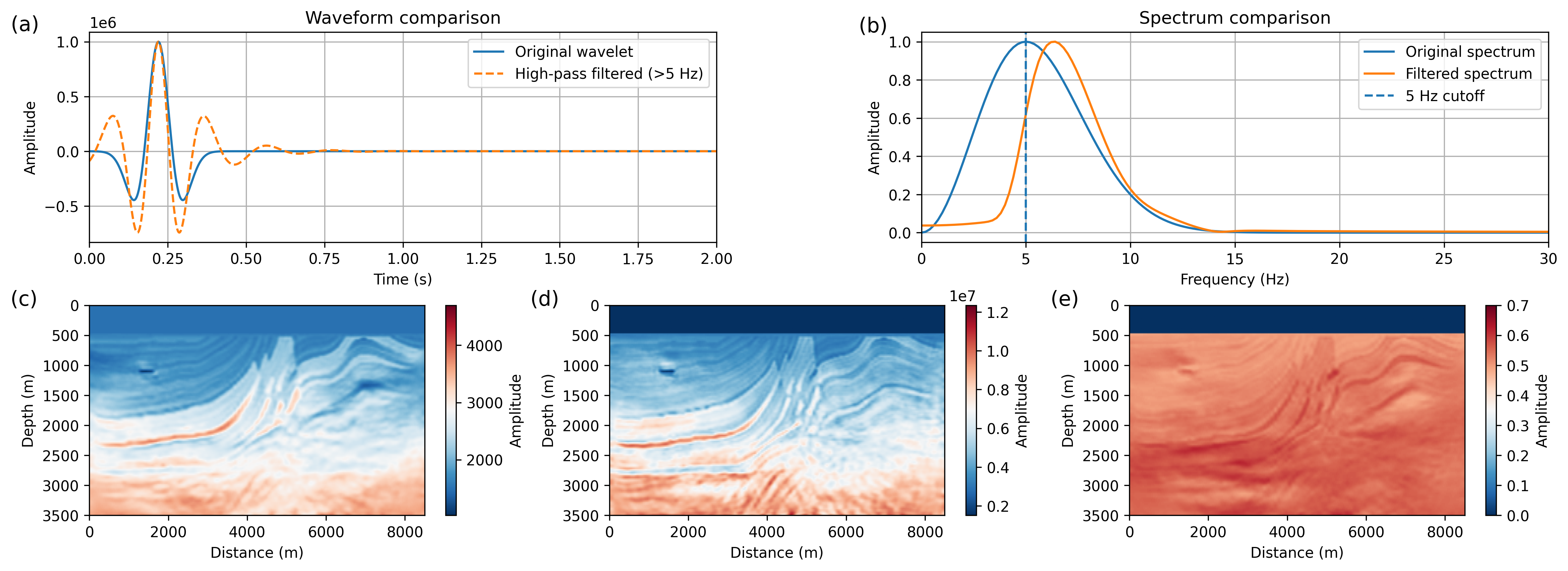}
\caption{(a) and (b) show the source and spectrum comparison of the original and filtered wavelet. (c)-(e) show inversion results using the filtered wavelet.}
\label{fig:lowfreq}
\end{figure*}

In this section, we investigate how missing low-frequency information affects the proposed INR-EFWI framework. Based on the Marmousi model shown in Figure~\ref{fig:marmini}, we adopt the same inversion settings as in the previous experiment, including the learning rates, number of iterations, acquisition geometry, and initial models. Figures~\ref{fig:lowfreq}(a) and~\ref{fig:lowfreq}(b) compare the original wavelet with the filtered wavelet and their corresponding normalized amplitude spectra. As shown in Figure~\ref{fig:lowfreq}(b), frequency components below 5 Hz are strongly suppressed, resulting in a more challenging inversion scenario with limited low-frequency information. Figures~\ref{fig:lowfreq}(c)-(e) present the corresponding inverted Vp, impedance, and Vs/Vp models obtained using INR-EFWI. Despite the loss of low-frequency information, the proposed method still reconstructs the major structural features of the model and preserves relatively good lateral continuity. The final MAE, MSNR, and SSIM values are summarized in Table~\ref{table2}. Compared with the inversion results obtained using the original wavelet in Figure~\ref{fig:marminv}, reconstruction accuracy degrades, as expected, because the filtered data has less low-wavenumber information. Nevertheless, INR-EFWI remains stable and can still recover the dominant subsurface structures, demonstrating improved robustness under a more challenging low-frequency-deficient condition.

\subsection{Sensitivity to noise level}
\begin{figure*}
\centering
\includegraphics[width=1.0\columnwidth]{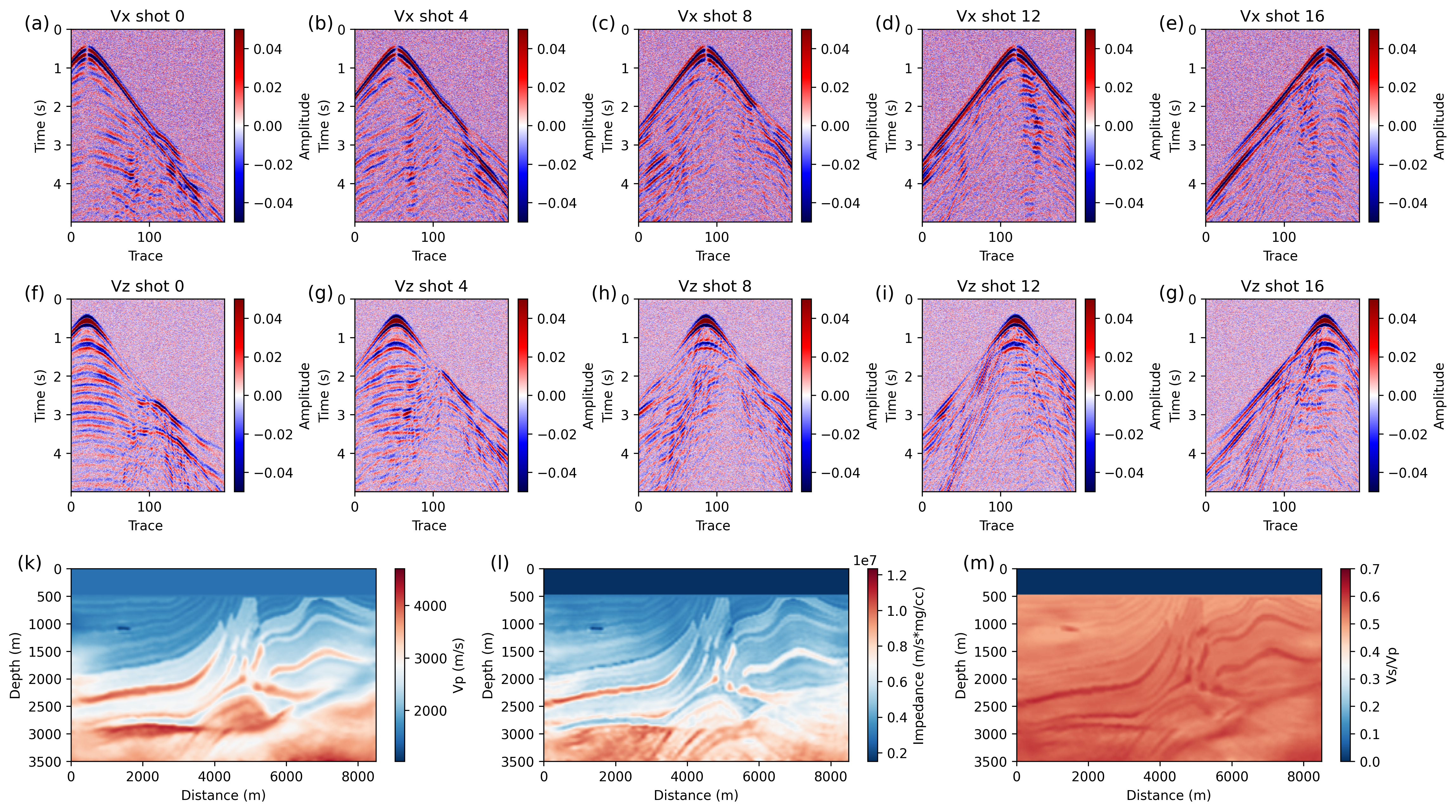}
\caption{(a)-(e) show noisy shot gathers of x component at different locations. (f)-(g) show noisy shot gathers of the z component at different locations. (k) The inverted Vp model using the noisy dataset. (l) The inverted impedance model using the noisy dataset. (m) The inverted Vs/Vp model using the noisy dataset.}
\label{fig:noise}
\end{figure*}

Next, we further evaluate the robustness of the proposed INR-EFWI framework in the presence of ambient noise. We keep all inversion settings unchanged and add Gaussian noise to the observed data to achieve a data signal-to-noise ratio (DSNR) of 1 dB. Figures~\ref{fig:noise}(a)-(j) show representative noisy $v_x$- and $v_z$-component shot gathers at different source locations. Owing to the strong noise contamination, a considerable portion of the weaker seismic events is partially obscured, making the inversion problem substantially more challenging. Using these noisy observations, we repeat the inversion experiment, and Figures~\ref{fig:noise}(k)-(m) present the resulting Vp, impedance, and Vs/Vp models, respectively. Table~\ref{table2} summarizes the corresponding quantitative metrics. Despite the noise contamination, the recovered models still preserve the main structural features and overall geological continuity of the ground-truth models, with only a limited degradation in reconstruction quality. These results show that the proposed INR-EFWI framework is robust and noise-resistant under low-DSNR conditions.

\begin{table*}[!t]%
\centering %
\caption{Quantitative comparison of INR-EFWI under different conditions. The symbols $\uparrow$ and $\downarrow$ indicate that larger and smaller values are better, respectively.
\label{table2}}%
\begin{tabular*}{\textwidth}{@{\extracolsep\fill}llccc@{\extracolsep\fill}}
\toprule
Condition
& Parameter
& {MAE $\downarrow$}
& {MSNR (dB) $\uparrow$}
& {SSIM $\uparrow$} \\

\midrule
\multirow{3}{*}{W/O low frequency}
& $V_p$
& 351.60
& 13.74
& 0.40 \\
& $I_p$
& 8.8$\times 10^5$
& 13.41
& 0.38\\
& $V_s/V_p$
& 0.013
& 27.94
& 1 \\
\midrule
\multirow{3}{*}{Noisy (DSNR=1 dB)}

& $V_p$
& 263.02
& 15.59
& 0.53 \\
& $I_p$
& 8.3$\times 10^5$
& 14.00
& 0.53 \\
& $V_s/V_p$
& 0.012
&28.07
& 1 \\
\bottomrule
\end{tabular*}
\end{table*}

\section{CONCLUSION}
We propose a multiscale implicit-neural-representation-assisted elastic full-waveform inversion framework, termed INR-EFWI, to improve the robustness of multiparameter elastic inversion. By representing different physical parameters with independent network branches, INR-EFWI provides parameter-specific nonlinear representations that adapt to each variable's distinct sensitivities and spatial characteristics, reducing parameter interference and mitigating cross-talk during inversion. The framework further stabilizes the optimization in the absence of low-frequency components, improving convergence from challenging initial models. The effectiveness of the proposed method is demonstrated through both synthetic and field-data examples. Compared with AD-EFWI, INR-EFWI produces inversion results with fewer artifacts, better-preserved structural continuity, and improved model fidelity, while also achieving better agreement with the observed seismic data. These results indicate that INR-EFWI provides a more robust and effective framework for multiparameter elastic full waveform inversion.

\section{Data Availability Statement}
The data that support the findings of this study are available upon request from the corresponding author.

\bibliographystyle{unsrt}
\bibliography{my}

@article{taran1984,
    author = {Tarantola, Albert},
    title = {Inversion of seismic reflection data in the acoustic approximation},
    journal = {Geophysics},
    volume = {49},
    number = {8},
    pages = {1259-1266},
    year = {1984},
    doi = {10.1190/1.1441754},
}

@article{virieus2009,
author = {J. Virieux and S. Operto},
title = {An overview of full-waveform inversion in exploration geophysics},
journal = {Geophysics},
volume = {74},
number = {6},
pages = {WCC1-WCC26},
year = {2009},
doi = {10.1190/1.3238367},
}

@article{operto2015,
    author = {Operto, S. and Miniussi, A. and Brossier, R. and Combe, L. and Métivier, L. and Monteiller, V. and Ribodetti, A. and Virieux, J.},
    title = {Efficient 3-{D} frequency-domain mono-parameter full-waveform inversion of ocean-bottom cable data: application to Valhall in the visco-acoustic vertical transverse isotropic approximation},
    journal = {Geophysical Journal International},
    volume = {202},
    number = {2},
    pages = {1362-1391},
    year = {2015},
    doi = {10.1093/gji/ggv226},
}

@article{operto2013,
    author = {Operto, S. and Gholami, Y. and Prieux, V. and Ribodetti, A. and Brossier, R. and Metivier, L. and Virieux, J.},
    title = {A guided tour of multiparameter full-waveform inversion with multicomponent data: {F}rom theory to practice},
    journal = {The Leading Edge},
    volume = {32},
    number = {9},
    pages = {1040-1054},
    year = {2013},
    month = {09},
    doi = {10.1190/tle32091040.1},
}

@article{borisov2017,
    author = {Borisov, Dmitry and Modrak, Ryan and Gao, Fuchun and Tromp, Jeroen},
    title = {3{D} elastic full-waveform inversion of surface waves in the presence of irregular topography using an envelope-based misfit function},
    journal = {Geophysics},
    volume = {83},
    number = {1},
    pages = {R1-R11},
    year = {2017},
    month = {11},
    doi = {10.1190/geo2017-0081.1}
}

@article{prieux2013,
    author = {Prieux, Vincent and Brossier, Romain and Operto, Stéphane and Virieux, Jean},
    title = {Multiparameter full waveform inversion of multicomponent ocean-bottom-cable data from the Valhall field. {P}art 1: imaging compressional wave speed, density and attenuation},
    journal = {Geophysical Journal International},
    volume = {194},
    number = {3},
    pages = {1640-1664},
    year = {2013},
    month = {09},
    doi = {10.1093/gji/ggt177}
}

@article{brossier2009,
    author = {Brossier, Romain and Operto, Stéphane and Virieux, Jean},
    title = {Seismic imaging of complex onshore structures by 2{D} elastic frequency-domain full-waveform inversion},
    journal = {Geophysics},
    volume = {74},
    number = {6},
    pages = {WCC105-WCC118},
    year = {2009},
    month = {12},
    doi = {10.1190/1.3215771}
}

@ARTICLE{sunh2022,
  author={Sun, Hongyu and Demanet, Laurent},
  journal={IEEE Transactions on Geoscience and Remote Sensing}, 
  title={Deep {L}earning for {L}ow-{F}requency {E}xtrapolation of {M}ulticomponent {D}ata in {E}lastic {FWI}}, 
  year={2022},
  volume={60},
  number={},
  pages={1-11},
  doi={10.1109/TGRS.2021.3135790}}

@article{panwy2018,
    author = {Pan, Wenyong and Innanen, Kristopher A and Geng, Yu},
    title = {Elastic full-waveform inversion and parametrization analysis applied to walk-away vertical seismic profile data for unconventional (heavy oil) reservoir characterization},
    journal = {Geophysical Journal International},
    volume = {213},
    number = {3},
    pages = {1934-1968},
    year = {2018},
    month = {06},
    doi = {10.1093/gji/ggy087}
}

@article{kohn2012,
    author = {Köhn, D. and De Nil, D. and Kurzmann, A. and Przebindowska, A. and Bohlen, T.},
    title = {On the influence of model parametrization in elastic full waveform tomography},
    journal = {Geophysical Journal International},
    volume = {191},
    number = {1},
    pages = {325-345},
    year = {2012},
    month = {10},
    doi = {10.1111/j.1365-246X.2012.05633.x}
}

@article{Modrak2016,
    author = {Modrak, Ryan and Tromp, Jeroen},
    title = {Seismic waveform inversion best practices: regional, global and exploration test cases},
    journal = {Geophysical Journal International},
    volume = {206},
    number = {3},
    pages = {1864-1889},
    year = {2016},
    month = {09},
    doi = {10.1093/gji/ggw202}
}

@article{LI2019106,
title = {Full waveform inversion for velocity and density with rock physical relationship constraints},
journal = {Journal of Applied Geophysics},
volume = {167},
pages = {106-117},
year = {2019},
issn = {0926-9851},
doi = {https://doi.org/10.1016/j.jappgeo.2019.04.005},
url = {https://www.sciencedirect.com/science/article/pii/S0926985118305652},
author = {Yi Li and Hanming Gu},
}

@article{wangyw2016,
    author = {Wang, Yuwei and Dong, Liangguo and Liu, Yuzhu and Yang, Jizhong},
    title = {2D frequency-domain elastic full-waveform inversion using the block-diagonal pseudo-{H}essian approximation},
    journal = {Geophysics},
    volume = {81},
    number = {5},
    pages = {R247-R259},
    year = {2016},
    month = {07},
    doi = {10.1190/geo2015-0678.1}
}

@article{hunagjp2025,
    author = {Huang, Jianping and Liu, Zhang and Jin, Kejie and Ba, Kailun and Liu, Yuhang and Kong, Linghang and Cui, Chao and Li, Chuang},
    title = {A time-domain multi-parameter elastic full waveform inversion with pseudo-Hessian preconditioning},
    journal = {Applied Geophysics},
    volume = {22},
    pages = {660-671},
    year = {2025},
    doi = {10.1007/s11770-022-0994-7}
}

@article{gebre2026,
   author = "Gebre, M.G. and Araújo, J.M. De and Bezerra, C.",
   title = "Structure-{G}uided {G}radient {P}reconditioning for {M}ulti-{P}arameter {E}lastic {F}ull-{W}aveform {I}nversion", 
   journal= "",
   year = "2026",
   volume = "2026",
   number = "1",
   pages = "1-5",
   doi = "https://doi.org/10.3997/2214-4609.202682042"
  }

@article{aragao2020,
    author = {Aragao, Odette and Sava, Paul},
    title = {Elastic full-waveform inversion with probabilistic petrophysical model constraints},
    journal = {Geophysics},
    volume = {85},
    number = {2},
    pages = {R101-R111},
    year = {2020},
    month = {02},
    doi = {10.1190/geo2019-0285.1}
}

@article{huqi2021,
    author = {Hu, Qi and Keating, Scott and Innanen, Kristopher A. and Chen, Huaizhen},
    title = {Direct updating of rock-physics properties using elastic full-waveform inversion},
    journal = {Geophysics},
    volume = {86},
    number = {3},
    pages = {MR117-MR132},
    year = {2021},
    month = {04},
    doi = {10.1190/geo2020-0199.1}
}

@article{kamath2017,
    author = {Kamath, Nishant and Tsvankin, Ilya and Naeini, Ehsan Zabihi},
    title = {Facies-constrained FWI: Toward application to reservoir characterization},
    journal = {The Leading Edge},
    volume = {36},
    number = {11},
    pages = {924-930},
    year = {2017},
    month = {11},
    doi = {10.1190/tle36110924.1}
}

@article{aghazade2026,
author = {Aghazade, Kamal and Gholami, Ali},
title = {Robust {A}coustic and {E}lastic {F}ull {W}aveform {I}nversion by {A}daptive {T}ikhonov-{TV} {R}egularization},
journal = {SIAM Journal on Imaging Sciences},
volume = {19},
number = {1},
pages = {480-518},
year = {2026},
doi = {10.1137/25M1748238}
}

@article{taufik2024,
author = {Taufik, Mohammad H. and Wang, Fu and Alkhalifah, Tariq},
title = {Learned {R}egularizations for {M}ulti-{P}arameter {E}lastic {F}ull {W}aveform {I}nversion {U}sing {D}iffusion {M}odels},
journal = {Journal of Geophysical Research: Machine Learning and Computation},
volume = {1},
number = {1},
pages = {e2024JH000125},
doi = {https://doi.org/10.1029/2024JH000125},
year = {2024}
}

@article{sun2023implicit,
  title={Implicit seismic full waveform inversion with deep neural representation},
  author={Sun, Jian and Innanen, Kristopher and Zhang, Tianze and Trad, Daniel},
  journal={Journal of Geophysical Research: Solid Earth},
  volume={128},
  number={3},
  pages={e2022JB025964},
  year={2023},
  publisher={Wiley Online Library}
}

@article{cheny2026,
    author = {Chen, Yue and Sun, Jian and Zhu, Zhaorui and Song, Peng and Tan, Jun},
    title = {Perturbation-{B}ased {I}mplicit {F}ull-{W}aveform {I}nversion with {I}mproved {O}ptimization {C}onditioning},
    journal = {Geophysical Journal International},
    pages = {ggag403},
    year = {2026},
    month = {09},
    doi = {10.1093/gji/ggag403}
}

@ARTICLE{chenrh2025,
  author={Chen, Ruihua and Wu, Bangyu and Li, Meng and Luo, Yisi},
  journal={IEEE Transactions on Geoscience and Remote Sensing}, 
  title={Full-{W}aveform {I}nversion {W}ith {V}elocity {M}odel {L}ow-{R}ank {I}mplicit {N}eural {R}epresentation}, 
  year={2025},
  volume={63},
  number={},
  pages={1-16},
  doi={10.1109/TGRS.2025.3594184}}

@article{kangb2026,
    author = {Kang, Bo and Chen, Ruihua and Yang, Kai and Li, Meng and Wu, Bangyu},
    title = {Implicit full waveform inversion with adaptive {F}ourier frequency bases learning},
    journal = {Geophysical Journal International},
    volume = {244},
    number = {1},
    pages = {ggaf404},
    year = {2026},
    month = {01},
    doi = {10.1093/gji/ggaf404}
}

@article{berit2026,
   author = "Berti, Sean and Aleardi, Mattia and Stucchi, Eusebio",
   title = "Implicit {N}eural {R}epresentation for {E}lastic {F}ull‐{W}aveform {I}nversion", 
   journal= "Geophysical Prospecting",
   year = "2026",
   volume = "74",
   number = "4",
   pages = "",
   doi = "https://doi.org/10.1111/1365-2478.70168"
  }

@article{wangzf2026,
author = {Wang, Zefeng and Cheng, Shijun and Mao, Weijian and Ouyang, Wei and Tang, Huanhuan},
title = {Meta-{L}earning-{E}nhanced {I}mplicit {F}ull {W}aveform {I}nversion},
journal = {Geophysical Prospecting},
volume = {74},
number = {7},
pages = {e70248},
doi = {https://doi.org/10.1111/1365-2478.70248},
year = {2026}
}

@article{pratt1998,
    author = {Pratt, R. Gerhard and Shin, Changsoo and Hick, G. J.},
    title = {Gauss–{N}ewton and full Newton methods in frequency–space seismic waveform inversion},
    journal = {Geophysical Journal International},
    volume = {133},
    number = {2},
    pages = {341-362},
    year = {1998},
    doi = {10.1046/j.1365-246X.1998.00498.x},
}

@article{licdlmefwi2026,
   author = "Li, Chao and Chen, Yangkang",
   title = "Deep Learning Matching Filtering for Elastic Full Waveform Inversion", 
   journal= "Geophysical Prospecting",
   year = "2026",
   volume = "74",
   number = "7",
   pages = "",
   doi = "10.1111/1365-2478.70236",
  }

@article{licdlmfwi2026,
    author = {Li, Chao and Fomel, Sergey and Chen, Yangkang},
    title = {D{LM}-{FWI}: deep learning matching filtering for full waveform inversion},
    journal = {Geophysical Journal International},
    volume = {245},
    number = {1},
    pages = {ggag019},
    year = {2026},
    month = {04},
    doi = {10.1093/gji/ggag019}
}

@article{barnier2023,
author = {Guillaume Barnier and Ettore Biondi and Robert G. Clapp and Biondo Biondi},
title = {Full-waveform inversion by model extension: {P}ractical applications},
journal = {Geophysics},
volume = {88},
number = {5},
pages = {R609-R643},
year = {2023},
doi = {10.1190/geo2022-0382.1},
}

@article{vigh2014,
author = {Denes Vigh and Kun Jiao and Dave Watts and Dong Sun},
title = {Elastic full-waveform inversion application using multicomponent measurements of seismic data collection},
journal = {Geophysics},
volume = {79},
number = {2},
pages = {R63-R77},
year = {2014},
doi = {10.1190/geo2013-0055.1}
}

@article{hicks2016,
    author = {Hicks, Erik and Hoeber, Henning and Houbiers, Marianne and Lescoffit, Séverine Pannetier and Ratcliffe, Andrew and Vinje, Vetle},
    title = {Time-lapse full-waveform inversion as a reservoir-monitoring tool — {A} {N}orth Sea case study},
    journal = {The Leading Edge},
    volume = {35},
    number = {10},
    pages = {850-858},
    year = {2016},
    doi = {10.1190/tle35100850.1},
}

@article{Licqawi2024,
    author = {Li, C and Liu, G and Li, F and Wang, Z},
    title = "{Robust joint adaptive multiparameter waveform inversion with attenuation compensation in viscoacoustic media}",
    journal = {Geophysics},
    pages = {R231-R246},
    year = {2024},
    month = {03},
    doi = {10.1190/geo2022-0663.1},
}

@article{crase1990,
    author = {Crase, E. and Pica, A. and Noble, M. and McDonald, J. and Tarantola, A.},
    title = {Robust elastic nonlinear waveform inversion; application to real data},
    journal = {Geophysics},
    volume = {55},
    number = {5},
    pages = {527-538},
    year = {1990},
    doi = {10.1190/1.1442864}
}

@article{singh2021elastic,
  title={Elastic {FWI} for orthorhombic media with lithologic constraints applied via machine learning},
  author={Singh, Sagar and Tsvankin, Ilya and Zabihi Naeini, Ehsan},
  journal={Geophysics},
  volume={86},
  number={4},
  pages={R589--R602},
  year={2021},
  publisher={Society of Exploration Geophysicists}
}

@article{wangsw2025,
   author = "Wang, S. and Ravasi, M. and Alkhalifah, T.",
   title = "Multiresolution {H}ash {E}ncoding for {H}igh {R}esolution {I}mplicit {F}ull {W}aveform {I}nversion", 
   journal= "",
   year = "2025",
   volume = "2025",
   number = "1",
   pages = "1-5",
   doi = "https://doi.org/10.3997/2214-4609.202510109",
  }

@article{sitzmann2020,
      title={Implicit {N}eural {R}epresentations with {P}eriodic {A}ctivation {F}unctions}, 
      author={Vincent Sitzmann and Julien N. P. Martel and Alexander W. Bergman and David B. Lindell and Gordon Wetzstein},
      year={2020},
      eprint={2006.09661},
      archivePrefix={arXiv},
      primaryClass={cs.CV},
      url={https://arxiv.org/abs/2006.09661}, 
}

@article{dhara2023elastic,
  title={Elastic full-waveform inversion using a physics-guided deep convolutional encoder--decoder},
  author={Dhara, Arnab and Sen, Mrinal K},
  journal={IEEE Transactions on Geoscience and Remote Sensing},
  volume={61},
  pages={1--18},
  year={2023},
  publisher={IEEE}
}

\end{document}